\documentclass[12pt]{article}

\usepackage[dvips]{graphicx}
\usepackage{epsfig}
\usepackage{amsmath,amsfonts,amssymb,amsthm}
\usepackage{mathrsfs,mathtools}
\usepackage{verbatim}
\usepackage{psfrag}
\usepackage{bm}
\usepackage{bbm}
\usepackage[utf8]{inputenc}
\usepackage[square,comma,sort&compress,numbers]{natbib}
\usepackage[dvipsnames]{xcolor}
\usepackage{slashed}
\usepackage{upgreek}
\usepackage{enumitem}
\usepackage{float}
\usepackage[T1]{fontenc}
\usepackage{soul}
\usepackage{pgfplots}
\usepackage{extarrows}
\pgfplotsset{compat=1.18}
\usetikzlibrary{external}

\usepackage[T1]{fontenc}

\tikzset{
  external/only named=true,
  thick/.style={line width=.5pt},
  approximation/.style={line width=1.2pt},
  numerics/.style={black, dotted, line width=.8pt},
  amplitude/.style={dashed},
  estimate/.style={dashed, line width=.8pt},
  normal plot/.style={line width=.8pt},
}

\usepackage[colorlinks=true,urlcolor=blue,anchorcolor=blue,citecolor=blue,filecolor=blue
,linkcolor=blue,menucolor=blue,linktocpage=true,pdfproducer=medialab,pdfa=true
]{hyperref}
\usepackage{epsf,epsfig}
\usepackage{graphics}
\usepackage{subcaption}

\newcommand\blfootnote[1]{%
  \begingroup
  \renewcommand\thefootnote{}\footnote{#1}%
  \addtocounter{footnote}{-1}%
  \endgroup
}

\def\d{\mathrm{d}}

\def\H{\mathcal{H}}

\def\O{\mathcal{O}}
\def\vec{\mathbf}
\def\i{\mathrm{i}}
\def\e{\mathrm{e}}

\def\T{\mathcal{T}}
\def\G{\mathcal{G}}

\def\M{\mathcal{M}}

\def\W{\mathcal{W}}

\usepackage[errorstop]{feynmp}
\newcommand{\rom}[1]{\uppercase\expandafter{\romannumeral #1\relax}}

\begin{document}

\thispagestyle{empty}

\begin{flushright}
{\small KCL-PH-TH/2023-07}
\end{flushright}

\vspace{0.4cm}

\begin{center}
\Large\bf\boldmath
False vacuum decay rates, more precisely
\end{center}

\vspace{-0.2cm}

\begin{center}
{Wen-Yuan Ai,*\blfootnote{*~wenyuan.ai@kcl.ac.uk} Jean Alexandre$^\dag$\blfootnote{$^\dag$~jean.alexandre@kcl.ac.uk} and Sarben Sarkar$^\ddag$\blfootnote{$^\ddag$~sarben.sarkar@kcl.ac.uk} \\
\vskip0.4cm
{\it Theoretical Particle Physics and Cosmology, King’s College London,\\ Strand, London WC2R 2LS, UK}\\
\vskip1.4cm}
\end{center}

\begin{abstract}
We develop a method for accurately calculating vacuum decay rates beyond the thin-wall regime in pure scalar field theory at the one-loop level of the effective action. It accounts for radiative effects resulting from quantum corrections to the classical bounce, including gradient effects stemming from the inhomogeneity of the bounce background. To achieve this, it is necessary to compute not only the functional determinant of the fluctuation operator in the background of the classical bounce but also its functional derivative evaluated at the classical bounce. The former is efficiently calculated using the Gel'fand-Yaglom method. We illustrate how the latter can also be calculated with the same method, combined with a computation of various Green's functions.
\end{abstract}

\newpage

\hrule
\tableofcontents
\vskip.85cm
\hrule


\section{Introduction}

One of the most striking features of the Standard Model (SM) is that the electroweak vacuum is metastable~\cite{Hung:1979dn,Lindner:1985uk,Sher:1988mj,Sher:1993mf,Casas:1994qy,Casas:1996aq,Isidori:2001bm,Burgess:2001tj,Isidori:2007vm,Ellis:2009tp,EliasMiro:2011aa}. The measured 125 GeV Higgs boson~\cite{Aad:2012tfa,Chatrchyan:2012ufa}, together with the 173 GeV top quark~\cite{Lancaster:2011wr}, suggests that the Higgs potential in the SM develops a lower minimum due to the renormalisation running of the Higgs self-interaction coupling. The electroweak vacuum could then decay to the true vacuum via tunnelling, which is a first-order phase transition in quantum field theory. State-of-the-art calculations suggest that the electroweak vacuum lies at the edge of the stable region, having a lifetime longer than the present age of the Universe~\cite{EliasMiro:2011aa,Degrassi:2012ry,Buttazzo:2013uya,DiLuzio:2015iua,Chigusa:2017dux,Andreassen:2017rzq}.\footnote{The scalar potential, the top Yukawa and the electroweak gauge couplings have been extracted from data with full 2-loop next-to-next-leading order (NNLO) precision~\cite{Degrassi:2012ry,Buttazzo:2013uya}. These parameters have been extrapolated to large energies with full 3-loop NNLO renormalisation-group-equation (RGE) precision~\cite{Buttazzo:2013uya}.}

In contrast to the running couplings, the radiative corrections to tunneling are computed less accurately. The transition rate is sensitive to a solitonic field configuration in Euclidean space, called the ``bounce''~\cite{Coleman:1977py,Callan:1977pt}. This bounce configuration also encodes the information of the nucleated critical bubble. Typically, the bounce is calculated from the tree-level scalar potential with the running coupling constants evaluated at the scale corresponding to the bounce radius. However, the bounce is an inhomogeneous configuration, and the beta functions for the couplings do not account for the effects of the background inhomogeneity. The inhomogeneity of the bounce requires us to go beyond the effective potential.
Such effects from background inhomogeneity, as well as the corresponding higher-loop radiative corrections, have not yet been included in the full SM computations but are only studied in simpler scalar \cite{Garbrecht:2015oea,Bezuglov:2018qpq,Bezuglov:2019uxg}, Yukawa \cite{Ai:2018guc}, and gauge~\cite{Ai:2020sru} theories in the planar thin-wall approximation.
These studies are based on a Green's function method developed earlier in Ref.~\cite{Baacke:2006kv}. See also Refs.~\cite{Bergner:2003au,Bergner:2003id,Baacke:2004xk} for the case of two-dimensional spacetime. 

We extend previous work by going beyond the thin-wall regime. In order to calculate the quantum-corrected bounce, we need both the functional determinants of the fluctuation operators and their functional derivatives evaluated at the classical bounce (see Eq.~\eqref{eq:Gamma-one-loop0} or Eq.~\eqref{eq:Gamma-one-loop} below). The functional determinants in the $O(4)$-symmetric backgrounds can be efficiently computed by the Gel'fand-Yaglom method~\cite{Gelfand:1959nq,Ai:2019fri}.  We show that the functional derivative of a functional determinant can be computed with the same method, combined with a computation of various Green's functions. This method is closely related to the Green's function method used in Refs.~\cite{Baacke:2006kv,Garbrecht:2015oea,Bezuglov:2018qpq,Ai:2018guc,Bezuglov:2019uxg,Ai:2020sru}. Our method is applied numerically to compute the corresponding radiative effects in an archetypal scalar model.

The outline of the paper is as follows. In the next section, we briefly review the standard Callan-Coleman formalism for false vacuum decay in terms of the classical bounce. In Sec.~\ref{sec:quantum-corrections}, we reformulate the decay rate using the one-particle-irreducible (1PI) effective action~\cite{Jackiw:1974cv} in a way that radiative corrections to the decay rate can be incorporated. In Sec.~\ref{sec:GY-method}, we introduce the powerful Gel'fand-Yaglom method for computing functional determinants. In Sec.~\ref{sec:function-derivative}, we show how the Gel'fand-Yaglom method, in combination with a Green's function method, is used to compute the quantum corrections to the classical bounce and the decay rate. In Sec.~\ref{sec:numerics} we give the results of numerical calculations using our method. In Sec.~\ref{sec:conc}, we give our conclusions. For completeness, some technical details are presented in the appendices. In Appendix~\ref{app:Green}, we give a general introduction to how to solve the Green's functions. In Appendix~\ref{app:regularisation}, we discuss how to renormalise the effective one-loop action using the dimensional renormalisation scheme. In the paper, we use $\hbar=c=1$.

\section{Callan--Coleman formalism and the tree-level bounce} \label{sec:CCformalism}

\begin{figure}[ht]
\centering
  \includegraphics[scale=0.6]{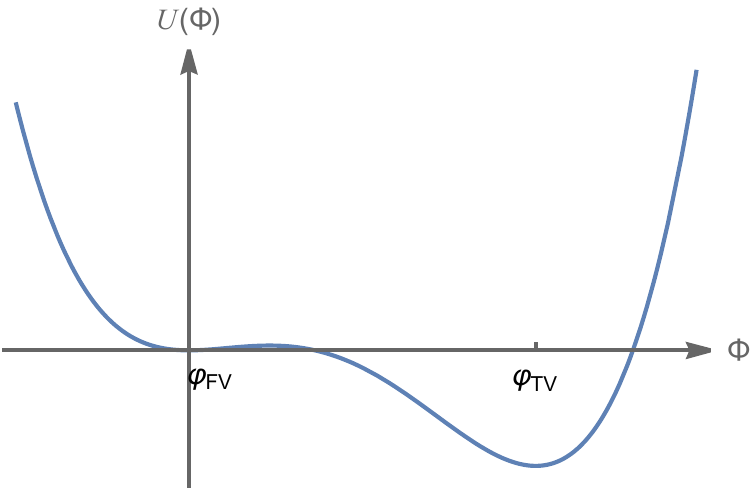}
\captionsetup{justification=raggedright,singlelinecheck=false}
  \caption{The classical  potential $U(\Phi)$ for the archetypical scalar model with false vacuum decay, given by Eqs.~\eqref{eq:themodel}
  and~\eqref{eq:classicalpotential}.}
  \label{fig:potential}
\end{figure}

We consider the archetypal model,
\begin{align}
\label{eq:themodel}
\mathcal{L}_{\rm M}=\frac{1}{2}\eta^{\mu\nu}(\partial_\mu\Phi)(\partial_\nu\Phi) - U(\Phi)\,,
\end{align}
where $\eta^{\mu\nu}$ is the Minkowski metric with signature $\{+,-,-,-\}$ and
\begin{align}
\label{eq:classicalpotential}
U(\Phi) = \frac{1}{2} m^2\Phi^2 - \frac{1}{3!}g\Phi^3 + \frac{1}{4!}\lambda\Phi^4 \,.
\end{align}
The couplings $m^2$, $g$, $\lambda$ all take positive values. When $g^2>8\lambda m^2/3$, the potential has two local minima. For $8\lambda m^2/3<g^2< 3\lambda m^2$, $\varphi=0$\footnote{We use $\varphi$ to denote the vacuum expectation value of the quantum field $\Phi$, $\varphi=\langle 0|\Phi|0\rangle$. This should be clearer when we introduce the effective action below.} is the global minimum; for  $g^2>3\lambda m^2$, the global minimum becomes
\begin{align}
     \varphi_{\rm TV}=\frac{3g+\sqrt{9g^2-24\lambda m^2}}{2\lambda}\,.
\end{align}
When $g^2=3\lambda m^2$, the two local minima become degenerate. Below we consider the parameter region $g^2>3\lambda m^2$. The potential (see Fig.~\ref{fig:potential}) has two minima at $\varphi_{\rm FV}=0$ and $\varphi_{\rm TV}$, corresponding to the false and true vacua respectively. In the potential, we have chosen $U(\varphi_{\rm FV})=0$ for convenience.

The decay rate is obtained from the Euclidean transition amplitude
\begin{align}
\label{eq:partition}
Z[0] = \langle \varphi_{\rm FV}| \e^{-H\mathcal{T}}|\varphi_{\rm FV}\rangle = \int\mathcal{D}\Phi\:\e^{-S[\Phi]}\,,
\end{align}
where $H$ is the Hamiltonian, $\mathcal{T}$ is the amount of Euclidean time
taken by the transition and $S$ is the Euclidean action. In terms of $Z[0]$, the decay rate per volume is given by~\cite{Callan:1977pt}
 \begin{align}
\frac{\gamma}{V}=\lim_{\T\rightarrow\infty}\frac{2}{V\T} |{\rm Im}(\ln Z[0])|\,.
\label{eq:decayrate2}
\end{align}

The partition function $Z[0]$ is
evaluated by using the method of steepest descent for path integrals. The classical equation of motion in Euclidean spacetime reads
\begin{align}
\label{eq:eom}
- \partial^2\varphi + U'(\varphi) = 0\,,
\end{align}
with the boundary conditions $\varphi|_{\tau\rightarrow\pm\infty} =
\varphi_{\rm FV}$ and $\dot{\varphi}|_{\tau=0}=0$, where the dot denotes the derivative with
respect to the Euclidean time $\tau$. A trivial solution is $\varphi(x)=\varphi_{\rm FV}$.
There is a nontrivial solution---the {\it bounce}---which is $O(4)$ symmetric in Euclidean spacetime. The bounce solution is relevant for false vacuum decay and describes a field configuration that starts and ends in Euclidean time in the false vacuum. Working in four-dimensional hyperspherical coordinates, the equation of motion reads
\begin{align}
\label{eq:eom-classical}
-\frac{\d^2\varphi(r)}{\d r^2} - \frac{3}{r} \frac{\d\varphi(r)}{\d r} + U'(\varphi(r)) = 0\,,
\end{align}
where $r=\sqrt{\tau^2+\vec{x}^2}$. The boundary conditions become $\varphi(r \rightarrow \infty)=\varphi_{\rm FV}$ and $\d\varphi(r)/\d r|_{r=0}=0$. We denote this classical bounce solution as $\varphi_b^{(0)}$.

Aside from the single bounce solution, there are also multi-bounce solutions that go back and forth from the false vacuum $N$ times, for arbitrary $N$. These multi-bounce configurations exponentiate~\cite{Callan:1977pt,Plascencia:2015pga,Ai:2020vhx} in analogous to disconnected vacuum Feynman diagrams. Integrating the fluctuations about the stationary points requires one to analyse the eigenfunctions of the fluctuation operators. In particular, the operator $-\partial^2+U''(\varphi^{(0)}_b)$ contains negative and zero modes (see Sec.~\ref{sec:GY-method} below). Carefully dealing with some subtleties from these particular modes, one obtains the decay rate per unit volume~\cite{Callan:1977pt}
\begin{align}
\label{eq:decay-oneloop}
\frac{\gamma}{V}=\left(\frac{B^{(0)}}{2\pi}\right)^2\left|\frac{\det^\prime[-\partial^2+U''(\varphi^{(0)}_b)]}{\det[-\partial^2+U''(\varphi_{\rm FV})]}\right|^{-1/2}\,\e^{-B^{(0)}}\,,
\end{align}
where $B^{(0)}=S[\varphi_b^{(0)}]-S[\varphi_{\rm FV}]$, $\det'$ means that the zero
eigenvalues are omitted from the determinant and a prefactor $\sqrt{B^{(0)}/2\pi}$ is included for each of the four collective coordinates that are related to the zero modes corresponding to spacetime translations~\cite{Gervais:1974dc}. Using $S[\varphi_{\rm FV}]=0$ we have
\begin{align}
\label{eq:B0}
    B^{(0)}=2\pi^2\int_0^\infty \d r\, r^3\left[\frac{1}{2}(\partial_r\varphi^{(0)}_b(r))^2+U(\varphi^{(0)}_b)\right]\,.
\end{align}

\section{Quantum-corrected bounce and radiative corrections to the decay rate}
\label{sec:quantum-corrections}

We reformulate Eq.~\eqref{eq:decayrate2} such that radiative effects can be systematically considered, using the 1PI effective
action~\cite{Jackiw:1974cv}. To keep the discussion general, we work in $d$-dimensional spacetime. But we will return to four-dimensional spacetime later. First, we introduce the partition function in the presence of a source $J$,
\begin{align}
Z[J] & = \int\mathcal{D}\Phi \,\e^{- S[\Phi] + \int\d^d x\, J(x)\Phi(x)}\,.
\label{eq:partitionfunctional}
\end{align}
The one-point function in the presence of the source $J$ is given by
\begin{align}
\varphi_{J}(x)& = \langle 0|\Phi(x)|0\rangle_{J} = \frac{\delta\ln
Z[J]}{\delta J(x)}\,.
\label{eq:onePointphi}
\end{align}
The effective
action then is defined as the Legendre transform
\begin{align}
\Gamma[\varphi_J]& = -\ln Z[J] +\int\d^d x\, J(x)\varphi_J(x)\,.
\label{eq:effectiveaction1}
\end{align}

From the effective action, one obtains the {\it quantum} equation of motion for the one-point function, $\delta\Gamma[\varphi_J]/\delta\varphi_J(x) = J(x)$. Note that $\Gamma$ has no explicit dependence on $J$. Denoting $\varphi\equiv \varphi_{J=0}$, one obtains the equation of motion for the vacuum expectation value in the absence of sources by setting $J=0$,
\begin{align}
\label{eq:quantumEoM}
\frac{\delta\Gamma[\varphi]}{\delta\varphi(x)} = 0\,.
\end{align}
A nontrivial $O(4)$-symmetric solution to Eq.~\eqref{eq:quantumEoM} would give the quantum-corrected bounce $\varphi_b$. In terms of the effective action, the decay rate per volume~\eqref{eq:decayrate2} can be
written as
\begin{align}
\label{eq:decay3}
\frac{\gamma}{V}= \frac{2\, |{\rm Im}\, \e^{-\left(\Gamma[\varphi_b]-\Gamma[\varphi_{\rm FV}]\right)}|}{V\mathcal{T}}\,.
\end{align}
This is the formula used in Refs.~\cite{Garbrecht:2015oea, Garbrecht:2015cla,Garbrecht:2015yza,
Plascencia:2015pga} where it is implicitly assumed that $\Gamma[\varphi_{\rm FV}]=0$. Since adding constant and linear terms in the Lagrangian does not affect the dynamics (when gravity is not considered), in the renormalisation procedure one can always implement $\varphi_{\rm FV}=0$ and $\Gamma[\varphi_{\rm FV}]=0$, even beyond the tree level.

On denoting the fluctuation operator evaluated at a general configuration $\varphi$ as
\begin{align}
\label{eq:def-M}
    \M [\varphi]=-\partial^2+U''(\varphi)\,,
\end{align}
the one-loop effective action reads
\begin{align}
\label{eq:Gammaphi1}
\Gamma^{(1)}[\varphi] = S[\varphi]+ \frac{1}{2}\,\ln\frac{\det \M[\varphi]}{\det \widehat{\M}} \equiv S[\varphi]+\frac{1}{2}D[\varphi]\,,
\end{align}
where $\widehat{\M}\equiv \M[\varphi_{\rm FV}]$ and $D[\varphi]\equiv\ln\frac{\det \M[\varphi]}{\det \widehat{\M}}$. Above, the normalisation choice $\Gamma[\varphi_{\rm FV}]=0$ is made.

The bounce at the one loop, $\varphi_b^{(1)}$, satisfies
\begin{align}
    \left.\frac{\delta S[\varphi]}{\delta\varphi(x)}\right|_{\varphi_b^{(1)}}+\frac{1}{2}\left.\frac{\delta D[\varphi]}{\delta\varphi(x)}\right|_{\varphi^{(1)}_b}=0\,.
\end{align}
Expanding this equation about $\varphi^{(0)}_b\equiv \varphi_b^{(1)}-\Delta\varphi_b$ gives\footnote{We assume that the one-loop corrected bounce is close to the classical bounce. If the false vacuum is generated by radiative corrections in the first place, it is necessary to use the two-particle irreducible effective action~\cite{Cornwall:1974vz} as
explained in Refs.~\cite{Garbrecht:2015cla,Garbrecht:2015yza}.}
\begin{align}
\label{eq:eom2}
[-\partial^2+U''(\varphi^{(0)}_b)]\Delta\varphi_b+\frac{1}{2}\left.\frac{\delta D[\varphi]}{\delta\varphi(x)}\right|_{\varphi^{(0)}_b}=0\,,
\end{align}
where we ignored perturbatively small higher-order terms. Substituting $\varphi^{(1)}_b=\varphi^{(0)}_b+\Delta\varphi_b$ into Eq.~\eqref{eq:Gammaphi1}, we obtain
\begin{align}
    \Gamma^{(1)}[\varphi^{(1)}_b]&= S[\varphi^{(0)}_b]+\frac{1}{2}D[\varphi^{(0)}_b]+\frac{1}{2}\int\d^d x\,  \Delta\varphi_b(x)\left[-\partial^2+U''(\varphi^{(0)}_b(x))\right]\Delta\varphi_b(x)\notag\\
    &+\frac{1}{2}\int\d^d x\,\left.\frac{\delta D[\varphi]}{\delta\varphi(x)}\right|_{\varphi_b^{(0)}}\Delta\varphi_b(x)\,,
\end{align}
Using Eq.~\eqref{eq:eom2}, we have 
\begin{align}
\label{eq:Gamma-one-loop0}
    \Gamma^{(1)}[\varphi^{(1)}_b]&= S[\varphi^{(0)}_b]+\frac{1}{2}D[\varphi^{(0)}_b]+\frac{1}{4}\int\d^d x\,\left.\frac{\delta D[\varphi]}{\delta\varphi(x)}\right|_{\varphi_b^{(0)}}\Delta\varphi_b(x)~.
\end{align}

When taking the functional derivative of $D[\varphi]$, we just focus on its $\varphi$ dependence and note that  $D[\varphi]\sim \ln\det \M[\varphi]={\rm tr}\ln \M[\varphi]$. In function space, we consider $\M[\varphi]$ as a ``diagonal matrix'' with one continuous index, $\M[\varphi(x)]$.\footnote{$\M$ can carry additional discrete indices for more general fields, e.g., spinor indices for spinor fields. In this case, we need to trace over the additional inner space.} Then 
\begin{align}
    \frac{\delta D[\varphi]}{\delta\varphi(x)}=\int \d^d x_1\d^d x_2 \M^{-1}[\varphi;x_1,x_2] \frac{\delta \M[\varphi(x_2)]}{\delta\varphi(x)}\,,
\end{align}
where the ``inverse matrix'' $\M^{-1}[\varphi;x_1,x_2]=G(x_1,x_2;\varphi)\delta^{(d)}(x_1-x_2)$ with $G(x_1,x_2;\varphi)$ being the Green's function to the operator $\M[\varphi]$. Therefore, we have 
\begin{align}
\label{eq:tad}
    \left.\frac{\delta D[\varphi]}{\delta\varphi(x)}\right|_{\varphi^{(0)}_{b}}=\M'[\varphi_b^{(0)}(x)]G(x,x;\varphi^{(0)}_{b})\,,
\end{align}
where a prime on $\M$ denotes the derivative with respect to the field. With this formula, one needs to calculate the Green's function $G(x,y;\varphi^{(0)}_{b})$ and take the coincident limit $x=y$. 

From Eq.~\eqref{eq:eom2}, $\Delta\varphi_b$ can be expressed as
\begin{align}
\label{eq:Deltavarphib-G}
    \Delta\varphi_b(x)=-\frac{1}{2}\int\d^d y\, G(x,y;\varphi_b^{(0)})\left.\frac{\delta D[\varphi]}{\delta\varphi(y)}\right|_{\varphi_b^{(0)}}=-\frac{1}{2}\int\d^d y\, G(x,y;\varphi_b^{(0)}) \M'[\varphi_b^{(0)}(y)]G(y,y;\varphi^{(0)}_{b}) \,, 
\end{align}
where we have used Eq.~\eqref{eq:tad}. Upon using $\M'[\varphi]=U'''(\varphi)=(-g+\lambda \varphi)$ (cf. Eq.~\eqref{eq:def-M}), the last term in Eq.~\eqref{eq:Gamma-one-loop0} reads
\begin{align}
    -\frac{1}{8}\int\d^d x\int\d^d y\, [-g+\lambda\varphi_b^{(0)}(x)] G(x,x;\varphi_b^{(0)}) G(x,y;\varphi_b^{(0)}) [-g+\lambda \varphi_b^{(0)}(y)]G(y,y;\varphi^{(0)}_{b})\,.
\end{align}
If one uses $G(x,y;\varphi_b^{(0)})$ as the propagator, the above expression can be pictorially represented as 
\begin{align}
\label{eq:Feynman}
\begin{tikzpicture}[baseline={-0.025cm*height("$=$")}]
\draw[draw=black,thick] (-0.5,0) circle (0.5);
\filldraw (0,0) circle (1.5pt) node {} ;
\draw[draw=black,thick] (0,0) -- (0.8,0);
\filldraw (0.8,0) circle (1.5pt) node {} ;
\draw[draw=black,thick] (1.3,0) circle (0.5);
\end{tikzpicture}
\,,\qquad
\begin{tikzpicture}[baseline={-0.025cm*height("$=$")}]
\draw[draw=black,thick] (-0.5,0) circle (0.5);
\filldraw (0,0) circle (1.5pt) node {} ;
\draw[draw=black,thick] (0,-0.6) circle (0.15) ;
\draw[draw=black,thick] (-0.5-0.12+0.5,0+0.12-0.6) -- (-0.5+0.12+0.5,0-0.12-0.6);
\draw[draw=black,thick] (-0.5-0.12+0.5,0-0.12-0.6) -- (-0.5+0.12+0.5,0+0.12-0.6);
\draw[draw=black,thick] (0,0) -- (0,-0.45);
\draw[draw=black,thick] (0,0) -- (0.8,0);
\filldraw (0.8,0) circle (1.5pt) node {} ;
\draw[draw=black,thick] (1.3,0) circle (0.5);
\end{tikzpicture}
\,,\qquad
\begin{tikzpicture}[baseline={-0.025cm*height("$=$")}]
\draw[draw=black,thick] (-0.5,0) circle (0.5);
\filldraw (0,0) circle (1.5pt) node {} ;
\draw[draw=black,thick] (0,-0.6) circle (0.15) ;
\draw[draw=black,thick] (-0.5-0.12+0.5,0+0.12-0.6) -- (-0.5+0.12+0.5,0-0.12-0.6);
\draw[draw=black,thick] (-0.5-0.12+0.5,0-0.12-0.6) -- (-0.5+0.12+0.5,0+0.12-0.6);
\draw[draw=black,thick] (0,0) -- (0,-0.45);
\draw[draw=black,thick] (0,0) -- (0.8,0);
\filldraw (0.8,0) circle (1.5pt) node {} ;
\draw[draw=black,thick] (1.3,0) circle (0.5);
\draw[draw=black,thick] (0.8,-0.6) circle (0.15) ;
\draw[draw=black,thick] (-0.5-0.12+1.3,0+0.12-0.6) -- (-0.5+0.12+1.3,0-0.12-0.6);
\draw[draw=black,thick] (-0.5-0.12+1.3,0-0.12-0.6) -- (-0.5+0.12+1.3,0+0.12-0.6);
\draw[draw=black,thick] (0.8,0) -- (0.8,-0.45);
\end{tikzpicture}\,,
\end{align}
where a  crossed wheel denotes the classical bounce $\varphi_b^{(0)}$. Note that this two-loop diagram is obtained from the expansion $\varphi_b^{(1)}=\varphi_b^{(0)}+\Delta\varphi_b$ of the {\it one-loop} diagram on using $G(x,y;\varphi_b^{(1)})$ as the propagator. When using $G(x,y;\varphi_b^{(0)})$ as the propagator, we say that the last term in Eq.~\eqref{eq:Gamma-one-loop0} corresponds to a \emph{partial} two-loop corrections, but not to full two-loop corrections.

\paragraph{Making use of  O(d)-symmetry.} Although $D[\varphi]$ and also the effective action are functionals of general configurations $\varphi(x)$, it is assumed  that the corrected bounce remains $O(d)$-symmetric. For this reason, it is sufficient to obtain $D[\varphi]$ for $O(d)$-symmetric configurations $\varphi(r)$ and take the functional derivative of the former with respect to the latter.
In this case, we can treat $\varphi(r)$ as a one-dimensional function and derive its equation of motion by directly varying the action $S[\varphi]$ and $D[\varphi]$ with respect to the one-dimensional field $\varphi(r)$. To illustrate this, we first consider the classical equation of motion for the bounce, \eqref{eq:eom-classical} generalised to  $d$-dimensional spacetime. As explained in Sec.~\ref{sec:CCformalism}, one can derive it from Eq.~\eqref{eq:eom} by assuming $O(d)$-symmetry. However, one can also derive it by applying the $O(d)$-symmetry to the classical action
\begin{align}
    S[\varphi]=\frac{2\pi^{d/2}}{\Gamma(d/2)}\int\d r\, r^{d-1} \left[\frac{1}{2}(\partial_r\varphi(r))^2+U(\varphi(r))\right]\,,
\end{align}
where $\Gamma(x)$ is the Gamma function, and varying it with respect to the one-dimensional field $\varphi(r)$.
Similarly, we can do the same procedure for $D[\varphi]$. Then the one-loop equation of motion reads
\begin{align}
    &\left.\frac{\delta S[\varphi]}{\delta\varphi(r)}\right|_{\varphi_b^{(1)}}+\frac{1}{2}\left.\frac{\delta D[\varphi]}{\delta\varphi(r)}\right|_{\varphi^{(1)}_b}=0 \notag\\
    &\qquad\qquad\qquad\Rightarrow  \frac{2\pi^{d/2} r^{d-1}}{\Gamma(d/2)} \left[-\frac{\d^2}{\d r^2}-\frac{d-1}{r}\frac{\d }{\d r}+U''(\varphi_b^{(0)})\right]\Delta\varphi_b(r)+\frac{1}{2}\left.\frac{\delta D[\varphi]}{\delta\varphi(r)}\right|_{\varphi^{(0)}_b}=0\,\label{E1},
\end{align}
where again we have expanded the functional derivative at the classical bounce and neglected perturbatively small higher-order terms.
Note that in this procedure $\delta D[\varphi]/\delta\varphi(r)|_{\varphi_b^{(0)}}$ is not equal to $\delta D[\varphi]/\delta\varphi(x)|_{\varphi_b^{(0)}}$; they even have different mass dimensions  due to the different Dirac delta functions generated in the two procedures for taking the functional derivative. The reason for introducing this new procedure is as follows. Below we have explicit expressions of $D[\varphi]$ only for $O(4)$-symmetric configurations $\varphi(r)$ instead of general four-dimensional configurations $\varphi(x)$. As a consequence, we can take the functional derivative of $D[\varphi]$ only with respect to $\varphi(r)$, rather than with respect to the general four-dimensional function $\varphi(x)$. Even when one assumes $O(d)$-symmetry for the one-point function, one still integrates over all possible fluctuations about $\varphi(r)$ in deriving the expression of $D[\varphi]$, as indicated by the sum over $l$ in Eqs.~\eqref{eq:renormalisedD1},~\eqref{eq:renormalisedD2} below.

From Eq.~\eqref{E1},
\begin{align}
\label{eq:Deltavarphib}
    \Delta\varphi_b(r)=-\frac{1}{2}\int\d r'\, \G(r,r')\left[\frac{\Gamma(d/2)}{2\pi^{d/2}r^{d-1}} \left.\frac{\delta D[\varphi]}{\delta\varphi(r')}\right|_{\varphi_b^{(0)}}\right]\,,
\end{align}
where $\G(r,r')$ is the Green's function, satisfying 
\begin{align}
\label{eq:Green-function-equation1}
    \left[-\frac{\d^2}{\d r^2}-\frac{d-1}{r}\frac{\d }{\d r}+U''(\varphi_b^{(0)})\right]\G(r,r')=\delta(r-r')\,.
\end{align}
This procedure finally gives
\begin{align}
\label{eq:Gamma-one-loop}
    \Gamma^{(1)}[\varphi^{(1)}_b]&= S[\varphi^{(0)}_b]+\frac{1}{2}D[\varphi^{(0)}_b]+\frac{1}{4}\int\d r\,\left.\frac{\delta D[\varphi]}{\delta\varphi(r)}\right|_{\varphi_b^{(0)}}\Delta\varphi_b(r)\,,
\end{align}
where the last term is a one-dimensional integral. Note that the last term above is \emph{not} obtained from the last term in Eq.~\eqref{eq:Gamma-one-loop0} by simply taking $\varphi(x)=\varphi(r)$. As we commented above, $D[\varphi]/\delta\varphi(r)|_{\varphi_b^{(0)}}\neq  \delta D[\varphi]/\delta\varphi(x)|_{\varphi_b^{(0)}}$ and we should have 
\begin{align}
    &\left.\left(\frac{1}{4}\int\d^d x\,\left.\frac{\delta D[\varphi]}{\delta\varphi(x)}\right|_{\varphi_b^{(0)}}\Delta\varphi_b(x)\right)\right|_{\varphi(x)=\varphi(r)}= \frac{1}{4}\int\d r\,\left.\frac{\delta D[\varphi]}{\delta\varphi(r)}\right|_{\varphi_b^{(0)}}\Delta\varphi_b(r)\notag\\
    &\qquad\qquad\qquad\Rightarrow \left.\frac{\delta D[\varphi]}{\delta\varphi(r)}\right|_{\varphi_b^{(0)}}= \frac{2\pi^{d/2}r^{d-1}}{\Gamma(d/2)} \left.\frac{\delta D[\varphi]}{\delta\varphi(x)}\right|_{\varphi_b^{(0)}}\,.
\end{align}

From Eqs.~\eqref{eq:Deltavarphib} and~\eqref{eq:Gamma-one-loop}, we see that, at the one-loop level of the effective action, a self-consistent bounce solution and the corrected decay rate are derived from $D[\varphi^{(0)}_b]$ and the functional derivative $\delta D[\varphi]/\delta\varphi(r)|_{\varphi_b^{(0)}}$ (or $\delta D[\varphi]/\delta\varphi(x)|_{\varphi_b^{(0)}}$ if one uses Eqs.~\eqref{eq:eom2} and~\eqref{eq:Gamma-one-loop0})).

In the following, we use the Gel'fand-Yaglom method to compute the functional determinant $D[\varphi]$ as well as its functional derivative.

\section{Functional determinants for an \texorpdfstring{$O(4)$}{TEXT} background and the Gel'fand-Yaglom method}

\label{sec:GY-method}

The powerful Gel'fand-Yaglom method~\cite{Gelfand:1959nq} is widely employed in calculations for
tunneling in theoretical as well as phenomenological studies, see e.g. Refs.~\cite{Isidori:2001bm,Dunne:2005rt,Andreassen:2017rzq,Chigusa:2017dux,Ai:2019fri,Guada:2020ihz,Ivanov:2022osf}. It is used in BubbleDet~\cite{Ekstedt:2023sqc}, a recent Python package that aims to compute functional determinants.

For a general $O(d)$-symmetric configuration $\varphi$ that approaches zero (the false vacuum $\varphi_{\rm FV}$) for $r\rightarrow \infty$, we have the decomposition
\begin{align}
    D[\varphi]=\sum_{l=0}^\infty{\rm deg}(l;d)\ln\left(\frac{{\det}\M_l[\varphi]}{{\det} \widehat{\M}_l}\right)\,,
\end{align}
where\footnote{The operators $\M_l[\varphi]$ are obtained when substituting $\Psi(r,\vec{\theta})=\psi_l(r) Y_l(\vec{\theta})$ (where $Y_l (\vec{\theta})$ are the hyperspherical harmonics) into the eigenequations for $\M[\varphi]$. It is the same for $\widehat{\M}_l$.} 
\begin{subequations}
\begin{align}
    \M_l[\varphi]&=-\frac{\d^2}{\d r^2}-\frac{d-1}{r}\frac{\d}{\d r}+\frac{l\left(l+d-2\right)}{r^2}+U''(\varphi(r))\,,\\
    \widehat{\M}_l&=-\frac{\d^2}{\d r^2}-\frac{d-1}{r}\frac{\d}{\d r}+\frac{l\left(l+d-2\right)}{r^2}+m^2\,,
\end{align}
\end{subequations}
and
\begin{align}
    {\rm deg}(l;d)=\frac{(d+2l-2)\Gamma(d+l-2) }{\Gamma(l+1)\Gamma(d-1)}\,.
\end{align}

For each $l$, one can calculate the functional determinant ratio using the Gel'fand-Yaglom theorem~\cite{Gelfand:1959nq}. Define the functions $\psi_l(r)$ and $\hat{\psi}_l(r)$ as the solutions to the following equations
\begin{align}
\label{eq:Ml-GY-equations}
    \M_l[\varphi]\psi_l(r;\varphi)=0\,,\qquad \widehat{\M}_l \hat{\psi}_l(r)=0\,,
\end{align}
with the same leading regular behavior at $r=0$. The Gel'fand-Yaglom theorem states that
\begin{align}
    \frac{{\det}\M_l[\varphi]}{{\det} \widehat{\M}_l}=\frac{\psi_l(\infty;\varphi)}{\hat{\psi}_l(\infty)}\equiv T_l(\infty;\varphi)\,.
\end{align}

First, we show an asymptotic analysis for $\hat{\psi}_l(r)$ and $\psi_l(r)$. For $r\rightarrow 0$, one can ignore the potential term in the operators $\widehat{\M}_l$ and $\M_l[\varphi]$. Therefore, the solutions regular at $r=0$ are proportional to $r^{l}$. Without loss of generality, both $\hat\psi_l(r)$ and $\psi_l(r)$ are normalised to $r^{l}$ as $r\rightarrow 0$. Then we obtain
\begin{align}
\label{eq:hat-psi-l}
    \hat{\psi}_l(r)=\left(\frac{2}{m}\right)^{l+d/2-1}\Gamma\left(l+\frac{d}{2}\right)\,r^{1-d/2}I_{l+d/2-1}(m r)\,.
\end{align}
$I_{l+d/2-1}(x)$ are the modified Bessel functions of the first kind, satisfying
\begin{align}
    \left[-\frac{\d^2}{\d r^2}-\frac{1}{r}\frac{\d}{\d r}-\left(m^2+\frac{(l+d/2-1)^2}{r^2}\right)\right] I_{l+d/2-1}(m r)=0\,.
\end{align}
$I_{l+d/2-1}(m r)\sim r^{l+d/2-1} (m/2)^{l+d/2-1}/\Gamma(l+d/2)$ for $r\rightarrow 0$. Substituting $\psi_l=T_l\hat\psi_l$ into Eq.~\eqref{eq:Ml-GY-equations}, one obtains
\begin{align}
\label{eq:T_l}
    \left[-\frac{\d^2}{\d r^2}-\left(2m\frac{I_{l+d/2-2}(m r)}{I_{l+d/2-1}(m r)}-\frac{2l+d-3}{r}\right)\frac{\d}{\d r} +W(\varphi(r))\right]T_l(r;\varphi)= 0\,,
\end{align}
where 
\begin{align}
    W(\varphi)=U''(\varphi)-m^2\,.
\end{align} 
The boundary conditions are 
\begin{align}
    T_l(0;\varphi)=1\,,\quad \left.\frac{\d T_l(r;\varphi)}{\d r}\right|_{r=0}=0\,.
\end{align}
We note that Eq.~\eqref{eq:T_l} can be written in a slightly different form due to the recursion relations of the modified Bessel functions.   

In the following, we take $d=4$.

\subsection*{Divergences and renormalisation}

The sum 
\begin{align}
\label{eq:sum}
    \sum_{l=0}^\infty (l+1)^2 \ln T_l(\infty;\varphi)
\end{align}
is divergent. This is not surprising because we are calculating the one-loop effective action whose divergence can only be removed by counterterms. Formally, we can write the renormalised one-loop effective action as
\begin{align}
\label{eq:formal-one-loop}
    S^{\rm ren}_1[\varphi]=S_0[\varphi]+\Delta S[\varphi] +S_{\rm ct}[\varphi]\,,
\end{align}
where $S_0[\varphi]$ is the classical action, $\Delta S[\varphi]= D[\varphi]/2$, and $S_{\rm ct}[\varphi]$ is the contribution from the counterterms. In the renormalised effective action, all the couplings (including mass parameters) are now renormalised ones. The details of the counterterms and renormalised couplings depend on the renormalisation scheme.. In the $\overline{\rm MS}$ scheme, we have 
\begin{align}
\label{eq:S1-MSbar}
    S^{\rm ren}_1[\varphi]=S_0^{\overline{\rm MS}}[\varphi]+\left(\Delta S[\varphi]-(\Delta S[\varphi])^{\rm pole}\right)\equiv S_0^{\overline{\rm MS}}[\varphi]+ (\Delta S[\varphi])^{\rm reg}\,,
\end{align}
where $S_0^{\overline{\rm MS}}$ is the lowest-order action expressed in terms of the renormalised $\overline{\rm MS}$ couplings, and $(\Delta S[\varphi])^{\rm pole}$ is the pole of the divergent part of $\Delta S$ defined according to the $\overline{\rm MS}$ renormalisation prescription. 

In the literature, two slightly different methods of obtaining a $(\Delta S[\varphi])^{\rm reg}$ are used. The first was used by Dunne and Min~\cite{Dunne:2005rt} (see also~\cite{Dunne:2006ct,Dunne:2007rt}), and followed by Refs.~\cite{Guada:2020ihz,Ivanov:2022osf,Ekstedt:2023sqc}. The second one was used even earlier by Baacke and Kiselev~\cite{Baacke:1993ne}, and followed in Refs.~\cite{Isidori:2001bm,Baacke:2003uw,Branchina:2014rva, Andreassen:2017rzq}. Below, we shall show that when the functional derivative with respect to $\varphi$ is taken, the first method gives a UV divergent result. This means that the first method is at most valid only ``on shell'', i.e., only for the classical bounce solution but not for variations about it. The second method still gives a UV-convergent functional derivative. However, we observe that the regularised one-loop functional determinant contains a UV-finite tadpole term (linear term in $\varphi$) for our model, which would shift the position of the false vacuum from zero. For self-consistency, we therefore propose removing such a non-vanishing tadpole by a linear counterterm. This would modify one of the finite terms ($A_{\rm fin}^{(1)}$) to be discussed below. 

Before we write down the explicit expressions for $(\Delta S[\varphi])^{\rm reg}$, we discuss why there are  two different points of view on the divergence in Eq.~\eqref{eq:sum}. The first one is that the divergence is caused by the large $l$ behaviour of $\ln T_l(\infty;\varphi)$ and one needs to derive the analytical expression of $\ln T_l(\infty;\varphi)$ as a series of $l$. This can be done by the WKB method for solving Eq.~\eqref{eq:T_l} for large $l$~\cite{Dunne:2005rt,Ekstedt:2023sqc}. Then one can use the dimensional regularisation to subtract the pole terms in the sum. This leads to the Dunne-Min formula given below. The second point of view is that the divergence is caused by $W(\varphi)\neq 0$, i.e., by a nonvanishing interaction term. If $W(\varphi)=0$, then the sum automatically gives zero. In this case, one expands $\ln T_l(\infty;\varphi)$ in powers of $W$. The divergence is found to be contained in the terms of order $\O(W)$, $\O(W^2)$ (each as a sum over $l$). These terms however correspond to the standard UV divergences of the effective action and thus can be expressed by loop integrals. (The inhomogeneity of the background field becomes irrelevant in the deep UV.) Once again, one can obtain pole terms in the loop integrals by dimensional regularisation which will be then subtracted from the sum. For more details see Appendix~\ref{app:regularisation}. This procedure leads to the Baacke-Kiselev formula, Eq.~\eqref{eq:renormalisedD2}.

The first method gives~\cite{Dunne:2005rt,Dunne:2006ct}
\begin{align}
\label{eq:renormalisedD1}
    &(\Delta S[\varphi])_{\rm DM}^{\rm reg}=\frac{1}{2}\sum_{l=0}^\infty (l+1)^2\left\{\ln T_l(\infty;\varphi)-\frac{\int_0^\infty\d r\, r W(\varphi)}{2(l+1)}+\frac{\int_0^\infty\d r\, r^3 W(\varphi)(W(\varphi)+2m^2)}{8(l+1)^3}\right\}\notag\\
    &-\frac{1}{16}\int_0^\infty\d r\, r^3 W(\varphi)\left(W(\varphi)+2m^2\right)\left[\log\left(\frac{\mu r }{2}\right)+\gamma_{\rm E}+1\right]\,,
\end{align}
where $\gamma_{\rm E}$ is Euler's constant, and $\mu$ is a renormalisation scale. 

The second method gives~\cite{Baacke:2003uw} 
\begin{align}
\label{eq:renormalisedD2}
    (\Delta S[\varphi])_{\rm BL}^{\rm reg}=&\frac{1}{2}\sum_{l=0}^\infty (l+1)^2 \left\{ \ln T_l(\infty;\varphi)-h_l^{(1)}(\infty;\varphi)-\left[h_l^{(2)}(\infty;\varphi)-\frac{1}{2} \left(h_l^{(1)}(\infty;\varphi)\right)^2\right]\right\}\notag\\
    &+\frac{1}{2}\left(A_{\rm fin}^{(1)}[\varphi]-\frac{1}{2}A_{\rm fin}^{(2)}[\varphi]\right)\,,
\end{align}
where $h_l^{(1)}$, $h_l^{(2)}$ satisfy
\begin{subequations}
\label{eq:h1l-h2l}
\begin{align}
\label{eq:hl1}
&\left[-\frac{\d^2}{\d r^2}-\left(2m\frac{I_{l}(mr)}{I_{l+1}(mr)}-\frac{2l+1}{r}\right)\frac{\d}{\d r}\right]h_l^{(1)}(r;\varphi)  +W(\varphi)=0\,,\\
&\left[-\frac{\d^2}{\d r^2}-\left(2m\frac{I_l(mr)}{I_{l+1}(mr)}-\frac{2l+1}{r}\right)\frac{\d}{\d r}\right]h_l^{(2)}(r;\varphi)+W(\varphi) h_l^{(1)}(r;\varphi)=0\,,
\end{align}
\end{subequations}
with boundary conditions $h_l^{(1,2)}(0;\varphi)=0, \d h_l^{(1,2)}(r;\varphi)/\d r|_{r=0}=0$\,,
and 
\begin{subequations}
\begin{align}
    A^{(1)}_{\rm fin}[\varphi]&=-\frac{m^2}{8}\left[1+\ln\left(\frac{\mu^2}{m^2}\right)\right]\int_0^\infty\d r\, r^3 W(\varphi(r))\,,\\
    A^{(2)}_{\rm fin}[\varphi]&=\frac{1}{8}\ln\left(\frac{\mu^2}{m^2}\right)\int\d r\, r^3 [W(\varphi(r))]^2\notag\\
    &+
    \frac{1}{16\pi^2} \int\frac{\d^4 k}{(2\pi)^4} |\widetilde{W}(k)|^2\left[2-\frac{\sqrt{k^2+4m^2}}{|k|}\ln\left(\frac{\sqrt{k^2+4m^2}+|k|}{\sqrt{k^2+4m^2}-|k|}\right)\right]\,,
\end{align}    
\end{subequations}
where 
\begin{align}
    \widetilde{W}(k)=\int \d^4 x\, \e^{\i k x} W(\varphi(x))\,.
\end{align}
In Appendix~\ref{app:regularisation}, we derive the above equations.
Since $\varphi(x)$ depends on $r=|x|$ only, we can write 
\begin{align}
\label{eq:tildeV}
    \widetilde{W}(k)=\frac{4\pi^2}{|k|} \int_0^\infty \d r\, r^2 W(\varphi(r)) J_1(|k| r)\equiv \widetilde{W}(|k|)\,,
\end{align}
where $J_1$ is the Bessel function of the first kind, defined as
\begin{align}
    J_1(|k| r)=\frac{|k| r}{\pi} \int_0^\pi \d\theta \,\sin^2\theta \,\e^{\i |k| r \cos\theta}\,.
\end{align}

We observe that $A_{\rm fin}^{(1)}$ contains a linear term in $\varphi$ for our model since $W(\varphi)=-g\varphi+\lambda\varphi^2/2$. As a consequence, 
\begin{align}
     \frac{\delta A_{\rm fin}^{(1)}[\varphi]}{\delta\varphi(r)}=-\frac{m^2}{8}\left[1+\ln\left(\frac{\mu^2}{m^2}\right)\right] r^3 U'''(\varphi(r))=\left[1+\ln\left(\frac{\mu^2}{m^2}\right)\right]\left(\frac{m^2 g }{8} r^3-\frac{m^2\lambda }{8}r^3\varphi(r)\right)\,,\label{eq:deltaA1-0}
\end{align}
is not zero even for $\varphi=0$. This means that the position of the false vacuum would be shifted if there are no other linear terms in Eq.~\eqref{eq:renormalisedD2}. We show below that such additional linear terms are absent. To be self-consistent, we have to add a linear counterterm to remove the observed tadpole term by requiring 
\begin{align}
    \left.\frac{\delta S_1^{\rm ren}[\varphi]}{\delta\varphi(x)}\right|_{\varphi=0}=0\,.
\end{align}
This is achieved by simply doing the following replacement,
\begin{align}
\label{eq:replacementA1}
    A_{\rm fin}^{(1)}[\varphi]\rightarrow \overline{A_{\rm fin}^{(1)}}[\varphi]=-\frac{m^2}{8}\left[1+\ln\left(\frac{\mu^2}{m^2}\right)\right]\int_0^\infty\d r\, r^3\, \overline{W(\varphi(r))}\,,
\end{align}
where an overline means the subtraction of tadpole terms. 

For false vacuum decay, it is natural to take $\mu=m$ (as we will do in the following) so that $\overline{A^{(1)}_{\rm fin}}[\varphi]$ and $A^{(2)}_{\rm fin}[\varphi]$ simplify further.

One can numerically solve Eq.~\eqref{eq:hl1} to obtain $h_l^{(1)}(r;\varphi)$ (and hence $h_l^{(1)}(\infty;\varphi)$), while $h_l^{(2)}(\infty;\varphi)$ as well as its functional derivative with respect to $\varphi$ can be computed via a Green's function method with known $h_l^{(1)}(r;\varphi)$, see below.

\subsubsection*{Zero and negative modes}

When taking $\varphi=\varphi^{(0)}_b$, there are four zero modes in the sector $l=1$. The
integral over the fluctuations in the zero-mode directions can be traded for an integral over the collective coordinates of the bounce, giving~\cite{Dunne:2006ct,Ekstedt:2023sqc} 
\begin{align}
\label{eq:zero-mode-contribution}
\e^{-\frac{1}{2}(1+1)^2 \ln T_1(\infty;\varphi^{(0)}_b)}\rightarrow   \underbrace{(V\T)\left(\frac{S_{\rm E}[\varphi_b^{(0)}]}{2\pi}\right)^{2}}_{\rm zero-mode\ contribution} \left(\frac{\det'\M_{1}[\varphi^{(0)}_b]}{\det\widehat{\M}_{1}}\right)^{-2} =(V\T)\left[2\pi A|\partial_r^2\varphi^{(0)}_b (0)|\right]^{2}\,,
\end{align}
where $A$ is defined by the normalisation of the asymptotic large $r$ behavior of the classical bounce: $\varphi_b^{(0)}(r)\sim A r^{-1} K_{1}(r)$. Here $K_1(r)$ is the modified Bessel function of the second kind. Using the equation of motion for $\varphi_b^{(0)}$, Eq.~\eqref{eq:eom-classical}, one has 
\begin{align}
    \left.\frac{\d^2 \varphi_b^{(0)}(r)}{\d r^2}\right|_{r=0}=\frac{U'(\varphi_b^{(0)}(0))}{4}\,,
\end{align}
where we have used $\partial_r\varphi_b^{(0)}(r)/r\rightarrow \partial^2_r\varphi_b^{(0)}(r)$ as $r\rightarrow 0$. So, formally, we have the replacement
\begin{align}
\label{eq:replacement-zero-mode}
   \ln T_1(\infty;\varphi_b^{(0)})\rightarrow -\frac{1}{2}\ln (V\T)-\frac{1}{2}\ln \left[(\pi A U'(\varphi_b^{(0)}(0)) /2)^2\right] \,.
\end{align}
Note that the factor of $V\T$ in Eq.~\eqref{eq:zero-mode-contribution} will be cancelled by the same factor in $\gamma/V$ (cf. Eq.~\eqref{eq:decay3}), leading to a finite decay rate per volume. 

The lowest eigenvalue mode in the $l = 0$ sector is a negative mode and is responsible for the instability of the false vacuum. Naively, the negative mode leads to divergence in the Gaussian functional integral about the bounce. However, this divergence is simply due to an improper application of the method of steepest descent~\cite{Callan:1977pt,Andreassen:2016cvx,Ai:2019fri}. With a suitable deformation of the contour into the complex plane whose real axis is generated by the negative mode, one actually obtains the following replacement
\begin{align}
\label{eq:negative-mode}
   T_0(\infty;\varphi_b^{(0)})\rightarrow\quad \left(\pm \frac{\i}{2}\right)^{-2} \left|T_0(\infty;\varphi_b^{(0)})\right|\,,
\end{align}
where the additional factor is due to the contour deformation and gives rise to an imaginary part of the effective action with a particular factor of $1/2$. The sign depends on the direction in which one deforms the contour into the complex plane. Usually, one chooses the way that leads to a positive decay rate. But in our formula~\eqref{eq:decay3}, we have taken the absolute value, and the sign in Eq.~\eqref{eq:negative-mode} does not matter. 
 
In the following, we show how to take the functional derivative in Eqs.~\eqref{eq:renormalisedD1} and~\eqref{eq:renormalisedD2}.

\section{Functional derivative of a functional determinant}
\label{sec:function-derivative}

To compute the quantum corrections to the bounce as well as the corresponding corrections to the decay rate, we still need to compute $\delta D[\varphi]/\delta\varphi(r)|_{\varphi^{(0)}_b}$ (cf. Eqs.~\eqref{eq:Deltavarphib} and~\eqref{eq:Gamma-one-loop}). We will use Eq.~\eqref{eq:renormalisedD2}, which leads
\begin{align}
\label{eq:delta-DeltaS}
    \frac{\delta (\Delta S[\varphi])_{\rm BL}^{\rm reg}}{\delta \varphi(r)}=&\frac{1}{2}\sum_{l=0}^{\infty} (l+1)^2\left\{\frac{1}{T_l(\infty;\varphi)}\frac{\delta T_l(\infty;\varphi)}{\delta\varphi(r)}-\frac{\delta h_l^{(1)}(\infty;\varphi)}{\delta\varphi(r)}-\frac{\delta h_l^{(2)}(\infty;\varphi)}{\delta\varphi(r)}\right.
    \notag \\
    &\qquad\qquad\qquad\left.+ h_l^{(1)}(\infty;\varphi)\frac{\delta h_l^{(1)}(\infty;\varphi)}{\delta\varphi(r)}\right\}+ \frac{1}{2}\frac{\delta \overline{A_{\rm fin}^{(1)}}(\varphi)}{\delta\varphi(r)}-\frac{1}{4} \frac{\delta A_{\rm fin}^{(1)}(\varphi)}{\delta\varphi(r)}\,.
\end{align}
At the end of this section, we shall comment on the problem if one uses Eq.~\eqref{eq:renormalisedD1}. 

We need to compute  $\delta T_l(\infty;\varphi)/\delta\varphi$, $\delta h_l^{(1,2)}(\infty;\varphi)/\delta \varphi$. To simplify the notation, we define
\begin{subequations}
    \begin{align}
     &\H_l[\varphi]\equiv \frac{\d^2}{\d r^2}+\left(2m\frac{I_{l}(m r)}{I_{l+1}(m r)}-\frac{2l+1}{r}\right)\frac{\d}{\d r} -W(\varphi(r)) \,,\\
     &\widehat{\H}_l\equiv \frac{\d^2}{\d r^2}+\left(2m\frac{I_{l}(m r)}{I_{l+1}(m r)}-\frac{2l+1}{r}\right)\frac{\d}{\d r}\,.
    \end{align}
\end{subequations}
Then 
\begin{align}
\label{eq:Hl-hatHl}
    \H_l[\varphi] T_l(r;\varphi)=0\,,\qquad \widehat{\H}_lh_l^{(1)}(r;\varphi)=W(\varphi(r))\,,\qquad \widehat{\H}_l h^{(2)}_l(r;\varphi)=W(\varphi(r)) h_l^{(1)}(r;\varphi)\,.
\end{align}

Taking the functional derivative of the first equation with respect to $\varphi(r)$ gives
\begin{align}
    \H_l[r';\varphi] \frac{\delta T_l(r';\varphi)}{\delta\varphi(r)}=-\frac{\delta \H_l [r';\varphi]}{\delta\varphi(r)} T_l(r';\varphi)=\delta(r'-r)U'''(\varphi(r')) T_l(r';\varphi)\,.
\end{align}
Viewing $r$ as a fixed number, the above equation in the coordinate $r'$ can then be solved by the Green's function method and we obtain
\begin{align}
    \frac{\delta T_l(r';\varphi)}{\delta\varphi(r)}&=\int\d r''\, G_l(r',r'';\varphi)\delta (r''-r)U'''(\varphi(r'')) T_l(r'';\varphi)=G_l(r',r;\varphi)U'''(\varphi(r))T_l(r;\varphi)\,,
\end{align}
where $G_l(r,r';\varphi)$ satisfies
\begin{align}
\label{eq:Gl-eq}
    \H_l[\varphi(r)]G_l(r,r';\varphi)=\delta(r-r')\,,
\end{align}
with the boundary conditions\footnote{To see these are the correct boundary conditions, one can analyse the equation $ \H_l[\varphi+\delta\varphi] T_l(r;\varphi+\delta\varphi)=0$ by writing $T_l(r;\varphi+\delta\varphi)=T_l(r;\varphi)+\delta T_l(r;\varphi)$. Since $T_l(r;\varphi)$ and $T_l(r;\varphi+\delta\varphi)$ satisfy the same boundary conditions imposed in the Gel'fand-Yaglom theorem, $\delta T_l(r;\varphi)$ has the boundary conditions $\delta T_l(0;\varphi)=0\,, \d\delta T_l(r;\varphi)/\d r|_{r=0}=0$ which are shared by $\delta T_l(r;\varphi)/\delta\varphi$ and thus the Green's function $G_l$. }
\begin{align}
\label{eq:Green-bdy}
    G_l(0,r';\varphi)=0\,,\quad \partial_r G_l(r,r';\varphi)|_{r=0}=0\,.
\end{align}
Taking $r'=\infty$ one obtains \begin{align}
\label{eq:central-formula}
  \frac{\delta T_l(\infty;\varphi)}{\delta \varphi(r)}=G_l(\infty,r;\varphi)U'''(\varphi(r)) T_l(r;\varphi)\,.
\end{align}

One can apply a similar analysis for the second and third equations in~\eqref{eq:Hl-hatHl}. But it is more straightforward to take the functional derivative with explicit expressions of $h_l^{(1,2)}(\infty;\varphi)$.  In this case, we need to solve the Green's function
\begin{align}
\label{eq:hatG-eq}
    \widehat{\H}_l\hat{G}_l(r,r')=\delta(r-r')\,,
\end{align}
with the same boundary conditions as in Eq.~\eqref{eq:Green-bdy}. Then 
\begin{subequations}
\label{eq:hl1-hl2-expressions}
    \begin{align}
        &h_l^{(l)}(\infty;\varphi)= \int\d r'\, \hat{G}_l(\infty,r') W(\varphi(r'))\,, \label{eq:hl1-expression}\\
        &h_l^{(2)}(\infty;\varphi)=\int\d r' \hat{G}_l(\infty,r') W(\varphi(r')) h_l^{(1)}(r';\varphi)\,. \label{eq:hl2-expression}
    \end{align}
\end{subequations}
The same expressions are valid if one replaces $\infty$ with $r$ above.
Taking the functional derivative of these expressions, one obtains
\begin{subequations}
\label{eq:hl1-hl2-ex}
    \begin{align}
        &\frac{\delta h_l^{(1)}(\infty;\varphi)}{\delta\varphi(r)}=\hat{G}_l(\infty,r) U'''(\varphi(r))\,,\\
        &\frac{\delta h_l^{(2)}(\infty;\varphi) }{\delta\varphi(r)}=\hat{G}_l(\infty,r)U'''(\varphi(r))h_l^{(1)}(r;\varphi)\notag\\
        &\qquad\qquad\qquad\qquad\qquad\qquad\qquad+\int \d r'\, \hat{G}_l(\infty,r') W(\varphi(r'))\hat{G}_l(r',r)U'''(\varphi(r))\,.
    \end{align}
\end{subequations}
$\hat{G}_l(r,r')$ can be obtained analytically (see Appendix~\ref{app:Green}) and reads
\begin{align}
\label{eq:hatG-analytic-result}
    \hat{G}_l(r,r')=\theta(r-r')  \frac{r' I_{l+1}(m r')}{I_{l+1}(mr)} \left[I_{l+1}(m r)K_{l+1}(m r')-I_{l+1}(mr')K_{l+1}(mr)\right]\,,
\end{align}
where $K_{l+1}(x)$ is the modified Bessel function of the second kind.
The Green's functions $G_l(r,r';\varphi)$ depend on $\varphi$ and can only be obtained numerically.  

From Appendix~\ref{app:Green}, we can see 
\begin{align}
\label{eq:general-property-Greensfunction}
G_l(r,\infty;\varphi)=\hat{G}_l(r,\infty)=0\,.
\end{align}
Therefore, one immediately has 
\begin{align}
    \left.\frac{\delta T_l(\infty;\varphi)}{\delta\varphi(r)}\right|_{r=\infty}= 0\,.
\end{align}
If one uses Eq.~\eqref{eq:renormalisedD1}, one would obtain a divergent {\it and cutoff-dependent} result from the second and third terms in the sum over $l$ at $r=\infty$. Therefore, Eq.~\eqref{eq:renormalisedD1} cannot give a meaningful result once the functional derivative with respect to $\varphi(r)$ is taken. On the contrary, Eq.~\eqref{eq:renormalisedD2} gives a finite result because we also have 
\begin{align}
   \left.\frac{\delta h_l^{(1)}(\infty;\varphi)}{\delta\varphi(r)}\right|_{r=\infty}=\left.\frac{\delta h_l^{(2)}(\infty;\varphi) }{\delta\varphi(r)}\right|_{r=\infty}=0\,,
\end{align}
due to the properties in Eq.~\eqref{eq:general-property-Greensfunction}. This also shows that there is no additional tadpole term in the sum over $l$ in Eq.~\eqref{eq:renormalisedD2}, otherwise one would obtain a term proportional to $r^3$ in the functional derivative, which is not vanishing for $r\rightarrow\infty$.

At last, the functional derivatives of $\overline{A^{(1)}_{\rm fin}}[\varphi]$ and $A^{(2)}_{\rm fin}[\varphi]$ with respect to $\varphi(r)$ read
\begin{subequations}
\label{eq:deltaAfin}
    \begin{align}
        \frac{\delta \overline{A_{\rm fin}^{(1)}}[\varphi]}{\delta\varphi(r)}=&-\frac{m^2\lambda }{8}r^3\varphi(r)\,,\label{eq:deltaA1}\\
        \frac{\delta A_{\rm fin}^{(2)}[\varphi]}{\delta\varphi(r)}=&\,\frac{r^2}{16\pi^2} U'''(\varphi(r))\int \d|k|\, |k|^2 J_1(|k|r) \widetilde{W}(|k|)\notag\\
        &\times \left[2-\frac{\sqrt{|k|^2+4m^2}}{|k|}\ln\left(\frac{\sqrt{|k|^2+4m^2}+|k|}{\sqrt{|k|^2+4m^2}-|k|}\right) \right]\,,
    \end{align}
\end{subequations}
where $\widetilde{W}(|k|)$ is defined in Eq.~\eqref{eq:tildeV}.

\section{Numerical example}
\label{sec:numerics}

With the analysis given in the previous sections, here we discuss the procedure for numerical calculations and provide a numerical example. We leave a more delicate phenomenological study for future work. 

Denoting
\begin{align}
\label{eq:B1-B2}
    \frac{1}{2} D[\varphi_b^{(0)}] =B^{(1)}-\ln\left(\pm \frac{\i}{2}\right)-\ln (V\mathcal{T})\,,\quad B^{(2)}=\frac{1}{4}\int\d r\, \left.\frac{\delta D[\varphi]}{\delta\varphi(r)}\right|_{\varphi_b^{(0)}}\Delta\varphi_b(r)\,,
\end{align}
we can write the decay rate per unit volume as
\begin{align}
    \frac{\gamma}{V}=\frac{2\,{\rm  Im}\,\e^{-\Gamma[\varphi_b^{(1)}]}}{V\T}=\e^{-B^{(0)}-B^{(1)}-B^{(2)}}\,.
\end{align}
Recall that $D[\varphi]/2=\Delta S[\varphi]$. Both here and below, $D[\varphi]$ is  regularised  even though the superscript ``reg'' is suppressed. We use the Baacke-Kiselev formula for the regularised one-loop functional determinant (cf. Eq.~\eqref{eq:renormalisedD2}) with the modification~\eqref{eq:replacementA1}.

In the numerical calculations, all the parameters that have mass dimension one take values in the unit of $m$ so that $m=1$. As a benchmark point, we choose 
\begin{align*}
    g=0.66\,,\qquad \lambda=0.1\,.
\end{align*}
The profile of the potential is shown in Fig.~\ref{fig:potential}.

\subsection{\texorpdfstring{$B^{(0)}$ and $ B^{(1)}$}{TEXT}}

For convenience of later discussion, we denote 
\begin{align}
    f_l[\varphi] \equiv \ln T_l(\infty;\varphi)-h_l^{(1)}(\infty;\varphi)-\left[h_l^{(2)}(\infty;\varphi)-\frac{1}{2} \left(h_l^{(1)}(\infty;\varphi)\right)^2\right]\,.
\end{align}
Then we have
\begin{align}
\label{eq:B1-specific}
    & B^{(1)}=\underbrace{\frac{1}{2}\ln\left|T_{0}(\infty;\varphi_b^{(0)})\right|}_{\text{negative-mode replacement~\eqref{eq:negative-mode}}}\underbrace{-\ln\left[(\pi A U'(\varphi_b^{(0)}(0)) /2)^2\right]}_{\text{zero-mode replacement~\eqref{eq:replacement-zero-mode}}}\underbrace{+\frac{1}{2}\sum_{l\geq 2}(l+1)^2 f_l[\varphi_b^{(0)}]}_{\text{positive-definite modes~\eqref{eq:renormalisedD2}}}\notag\\
    &\underbrace{-\frac{1}{2}\sum_{l=0,1} \left[h_l^{(1)}(\infty;\varphi_b^{(0)}) +h_l^{(2)}(\infty;\varphi_b^{(0)})-\frac{1}{2} \left(h_l^{(1)}(\infty;\varphi_b^{(0)})\right)^2\right] +\frac{1}{2}\left(\overline{A_{\rm fin}^{(1)}}[\varphi_b^{(0)}]-\frac{1}{2}A_{\rm fin}^{(2)}[\varphi_b^{(0)}]\right)}_{\text{remaining contribution generated from renormalisation, cf. Eq.~\eqref{eq:renormalisedD2},~\eqref{eq:replacementA1}}}\,.
\end{align}

We summarise the procedure of computing the bounce and decay rate at the level of the Callan-Coleman formula.

\begin{itemize}
    \item Step 1: Solve the classical EoM~\eqref{eq:eom-classical} and obtain the classical bounce $\varphi_b^{(0)}$. Substitute it into Eq.~\eqref{eq:B0} to obtain the classical bounce action $B^{(0)}$. 
    
    \item Step 2: Substitute $\varphi_b^{(0)}$ into Eq.~\eqref{eq:T_l}  to obtain $T_l(r;\varphi_b^{(0)})$, in particular its limiting value at $r\rightarrow\infty$, $T_l(\infty;\varphi_b^{(0)})$. 
    
    \item Step 3: Substituting $\hat{G}_l(\infty,r')$ (cf. Eq.~\eqref{eq:hatG-analytic-result}) and $\varphi_b^{(0)}$ into Eq.~\eqref{eq:hl1-expression} to obtain $h_l^{(1)}(r;\varphi_b^{(0)})$ and $h_l^{(1)}(\infty;\varphi_b^{(0)})$. Substituting $h_l^{(1)}(r;\varphi_b^{(0)})$ into Eq.~\eqref{eq:hl2-expression}, one obtains $h_l^{(2)}(\infty;\varphi_b^{(0)})$. 
\end{itemize}
The zero and negative modes in computing $T_l(\infty;\varphi_b^{(0)})$ need to be properly dealt with as discussed in Sec.~\ref{sec:GY-method}.
With these three steps, one can obtain $B^{(0)}$ and $B^{(1)}$ and thus the Callan-Coleman decay rate. 

\begin{figure}[ht]
    \centering
    \includegraphics[scale=0.6]{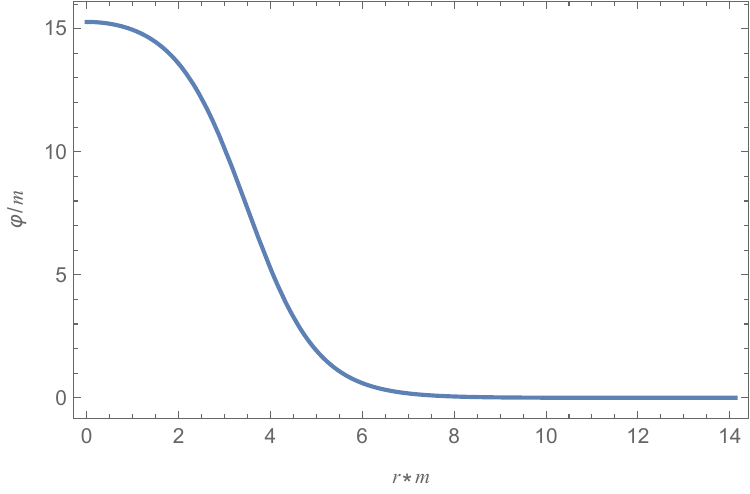}
    \caption{The profile of the classical bounce $\varphi_b^{(0)}(r)$.}
    \label{fig:ClassicalBounce}
\end{figure}

The classical bounce is obtained via a shooting algorithm and is shown in Fig.~\ref{fig:ClassicalBounce}.  The classical bounce action is $B^{(0)}=12942.7$.

\begin{figure}[ht]
    \centering
    \includegraphics[scale=0.6]{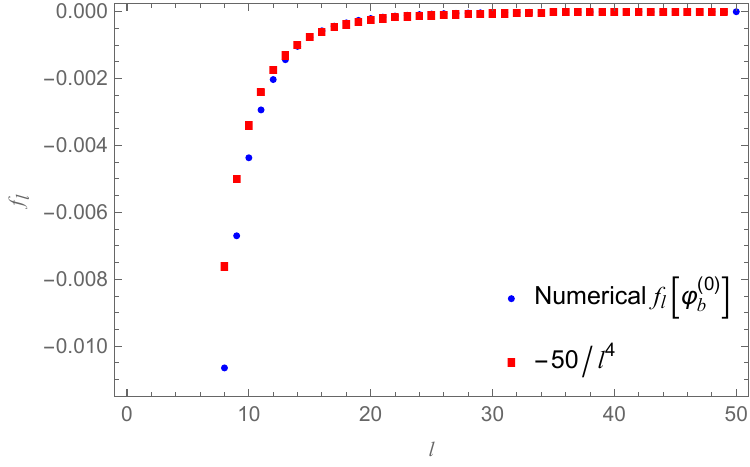}
    \caption{The behavior of $f_l$ as a function of $l$.}
    \label{fig:fl}
\end{figure}

For $B^{(1)}$, we first compute $f_l[\varphi_b^{(0)}]$
for $l\geq 2$. We find that $l_{\rm max}=50$ is already a good cutoff. We plot $f_l[\varphi_b^{(0)}]$ in Fig.~\ref{fig:fl} where we compare the numerical result with $-50/l^4$. Clearly $|f_l[\varphi_b^{(0)}]|$ drops as fast as or faster than $1/l^4$, ensuring a convergent result for the sum in Eq.~\eqref{eq:B1-specific}. We obtain
\begin{align}
    \frac{1}{2} \sum_{l\geq 2} (l+1)^2 f_l[\varphi_b^{(0)}] = -10.981\,. 
\end{align}
For the contribution from $l=0,1$, we find
\begin{subequations}
\begin{align}
     & B^{(1)}\supset  \frac{1}{2}\ln\left|T_{0}(\infty;\varphi_b^{(0)})\right|=-1.449\,,\\
    &B^{(1)}\supset -\ln\left[(\pi A U'(\varphi_b^{(0)}(0)) /2)^2\right]=-7.388\,,\\
    & B^{(1)}\supset -\frac{1}{2}\sum_{l=0,1} \left[h_l^{(1)}(\infty;\varphi_b^{(0)}) +h_l^{(2)}(\infty;\varphi_b^{(0)})-\frac{1}{2} \left(h_l^{(1)}(\infty;\varphi_b^{(0)})\right)^2\right]=-4.671\,.
\end{align}
\end{subequations}
Finally, $\overline{A_{\rm fin}^{(1)}}[\varphi_b^{(0)}]$ and $A_{\rm fin}^{(2)}[\varphi_b^{(0)}]$ are obtained as
\begin{align}
    \overline{A_{\rm fin}^{(1)}}[\varphi_b^{(0)}]=-15.177\,,\   A_{\rm fin}^{(2)}[\varphi_b^{(0)}]=-5.322 \quad \Rightarrow\quad  \frac{1}{2}\left(\overline{A_{\rm fin}^{(1)}}[\varphi_b^{(0)}]-\frac{1}{2}A_{\rm fin}^{(2)}[\varphi_b^{(0)}]\right)=-6.258\,.
\end{align}
In total, we have $B^{(1)}=-30.747$.

\subsection{\texorpdfstring{$B^{(2)}$}{TEXT}}

To obtain the quantum-corrected bounce and the decay rate at the full one-loop level of the effective action, we need to calculate $\delta D[\varphi]/\delta\varphi(r)|_{\varphi_b^{(0)}}$ (using Eq.~\eqref{eq:delta-DeltaS}) and $\Delta\varphi_b(r)$ (using Eq.~\eqref{eq:Deltavarphib}). The last two terms in Eq.~\eqref{eq:delta-DeltaS} are trivial because we have their analytical expressions (cf. Eqs.~\eqref{eq:deltaAfin}). The remaining calculation requires the following three further steps.

\begin{figure}
    \centering
    \includegraphics[scale=0.6]{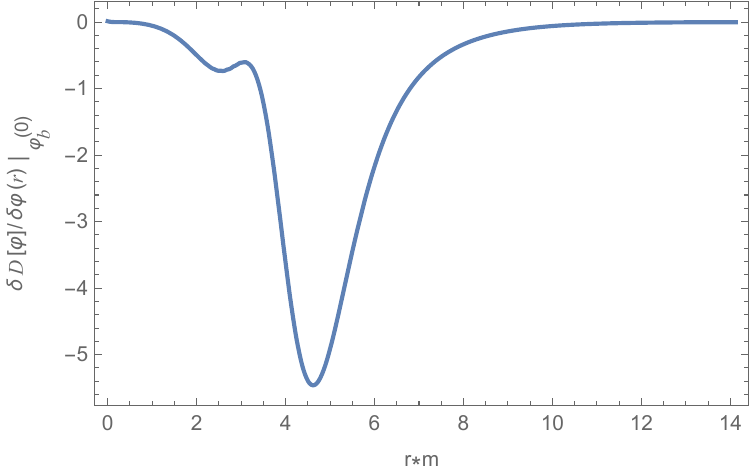}
    \caption{Numerical result for $\delta D[\varphi]/\delta\varphi(r)|_{\varphi_b^{(0)}}$.}
    \label{fig:deltaDdeltavarphi}
\end{figure}

\begin{itemize} 
    \item Step 4: Substituting $\varphi_b^{(0)}$ into Eq.~\eqref{eq:Gl-eq}, one can obtain $G_l(\infty,r';\varphi_b^{(0)})$. With $G_l(\infty,r';\varphi_b^{(0)})$ and $T_l(r;\varphi_b^{(0)})$,  one obtains $\delta T_l(\infty;\varphi)/\delta\varphi(r)|_{\varphi_b^{(0)}}$ through Eq.~\eqref{eq:central-formula}.

    Further, with $\hat{G}_l(r,r')$ and $h_l^{(1)}(r;\varphi_b^{(0)})$, one can obtain $\delta h_l^{(1)}(\infty;\varphi)/\delta\varphi(r)|_{\varphi_b^{(0)}}$ and\\ $\delta h_l^{(2)}(\infty;\varphi)/\delta\varphi(r)|_{\varphi_b^{(0)}}$ through Eqs.~\eqref{eq:hl1-hl2-ex}. 

    Substituting all of them and Eqs.~\eqref{eq:deltaAfin} into Eq.~\eqref{eq:delta-DeltaS}, one can finally obtain $\delta D[\varphi]/\delta\varphi(r)|_{\varphi_b^{(0)}}$.
    
    \item Step 5: Compute the Green's function in Eq.~\eqref{eq:Green-function-equation1} and substitute the obtained Green's function $\G(r,r')$ and $\delta D[\varphi]/\delta\varphi(r)|_{\varphi_b^{(0)}}$ into Eq.~\eqref{eq:Deltavarphib} to obtain $\Delta\varphi_b(r)$.
    
    \item Step 6: Finally, substitute the obtained $\Delta\varphi_b(r)$ and $\delta D[\varphi]/\delta\varphi(r)|_{\varphi_b^{(0)}}$ into the second equation in~\eqref{eq:B1-B2} to obtain $B^{(2)}$.
\end{itemize}
Again, in calculating the functional derivatives with respect to $\varphi$, the zero and negative modes need special treatment. {\it Formally}, 
\begin{align}
\label{eq:formal-function-derivative}
    \left.\frac{\delta D[\varphi]}{\delta\varphi(r)}\right|_{\varphi_b^{(0)}}=\lim_{\delta\varphi\rightarrow {\bf 0}}\frac{D[\varphi_b^{(0)}+\delta\varphi]-D[\varphi_b^{(0)}]}{\delta\varphi(r)}\,,
\end{align}
where ${\bf 0}$ is a function that equals to zero identically. In the above equation, one therefore needs to look into the spectrum for both $-\partial^2+U''(\varphi_b^{(0)})$ and $-\partial^2+U''(\varphi_b^{(0)}+\delta\varphi)$. We expect that for an infinitesimal deformation of the classical bounce $\delta\varphi\rightarrow {\bf 0}$, the operator $-\partial^2+U''(\varphi_b^{(0)}+\delta\varphi)$ still contains the same number of zero and negative modes. And we assume that Eq.~\eqref{eq:negative-mode} is still valid for $\varphi$ infinitely close to $\varphi_b^{(0)}$, so that one can obtain $\delta T_{0}(\infty;\varphi)/\delta\varphi(r)|_{\varphi_b^{(0)}}$ from that replacement. For the zero mode, it is more subtle because Eq.~\eqref{eq:replacement-zero-mode} is expressed in terms of $\varphi_b^{(0)}$ at $r=0$ and is not a standard functional of $\varphi_b^{(0)}(r)$. One even cannot formally replace $\varphi_b^{(0)}$ with $\varphi$ and take the functional derivative of the second term in Eq.~\eqref{eq:replacement-zero-mode}.
We thus will neglect the zero-mode contribution to the functional derivative.\footnote{The most important contribution of the zero modes to $D[\varphi]$ is the term involving the spacetime volume $V\T$. Contribution from this term is cancelled in Eq.~\eqref{eq:formal-function-derivative} and the remaining contribution is expected to be negligible. 
} However, we have checked that $
\delta h_{1}^{(1,2)}/\delta\varphi(r)|_{\varphi_b^{(0)}}$ are indeed negligible in the final result of $\delta D[\varphi]/\delta\varphi(r)|_{\varphi_b^{(0)}}$. The obtained $\delta D[\varphi]/\delta\varphi|_{\varphi_b^{(0)}}$ is shown in Fig.~\ref{fig:deltaDdeltavarphi}.

\begin{figure}[ht]
    \centering
    \includegraphics[scale=0.6]{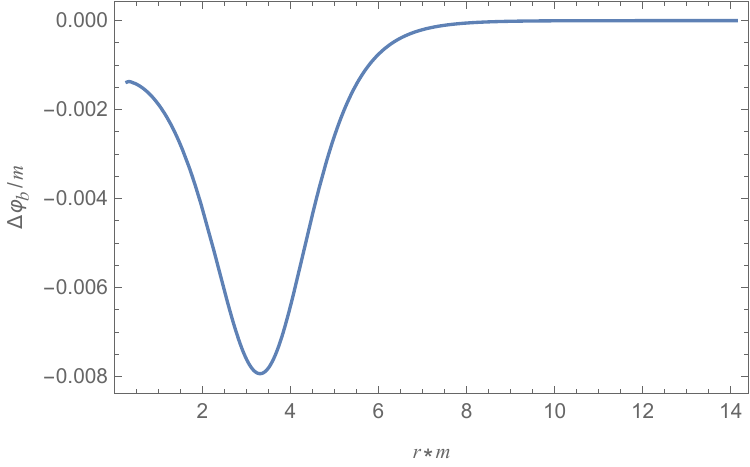} \quad
    \caption{The correction to the classical bounce, $\Delta\varphi(r)$.}
    \label{fig:BounceCorrection}
\end{figure}

Following Step 5, we obtain the correction to the classical bounce $\Delta\varphi_b(r)$ as shown Fig.~\ref{fig:BounceCorrection}. The profile of $\Delta\varphi_b(r)$ is negligible compared with $\varphi_b^{(0)}(r)$. However, the additional correction to the decay rate is obtained as $B^{(2)}=-16.42$ (Step 6), which is $53.4\%$ of $B^{(1)}$. The correction $\Delta\varphi_b(r)$ to the classical bounce is maximal around $rm\simeq 3.5$, which also corresponds to roughly the steepest part of the bubble wall profile. 
This indicates that these higher-order corrections receive large contributions from the gradients of the bubble wall.  This observation is consistent with that made in Ref.~\cite{Ai:2018guc} for thin-wall bubbles.

\section{Conclusions}
\label{sec:conc}
FVD is a very important phenomenon in quantum field theory and there continue to be many developments both conceptual and methodological~\cite{Ai:2019fri,Espinosa:2018hue,Braden:2018tky,Espinosa:2018voj,Espinosa:2018szu,Ai:2018rnh,Hertzberg:2019wgx,Chigusa:2019wxb,Espinosa:2019hbm,Darme:2019ubo,Mou:2019gyl,Croon:2021vtc,Hayashi:2021kro,Draper:2023fkz,Nishimura:2023dky,Blum:2023wnb,Batini:2023zpi,Steingasser:2023gde,Pirvu:2023plk}.
In this paper, we propose a systematic procedure to compute the self-consistent bounce solution and its resulting radiative correction to the decay rate at the one-loop level of the effective action. This is achieved by evaluating the one-loop effective action at the one-loop saddle point configuration for the bounce (the self-consistent bounce), instead of the classical one. The method requires the computation of both conventional functional determinants and their functional derivative, both evaluated at the classical bounce. Earlier studies on this are restricted to the thin-wall regime~\cite{Garbrecht:2015oea,Ai:2018guc,Ai:2020sru} and are usually based on the ``resolvent method'' of calculating functional determinants by treating Green's function as a spectral sum~\cite{Baacke:1993jr,Baacke:1993aj,Baacke:1994ix,Baacke:2008zx,Ai:2019fri}. In this work, we have demonstrated that the powerful Gel'fand-Yaglom method can be used to compute both the functional determinants and their functional derivatives, making the computation extremely efficient. In particular, for $O(4)$-symmetric configurations that asymptotically approach the false vacuum, an explicit expression for the renormalised one-loop effective action in pure scalar field theory is known. We have shown how it is directly used in the computation. 

Although the theoretical developments here are general, we provide an explicit numerical study of our archetypal model. Our chosen benchmark point for the mass and coupling parameters is such that the bounce is far away from the thin-wall regime. In this example, the correction $B^{(1)}$ (the decay rate due to the functional determinant evaluated at the classical bounce)  is small compared with the classical bounce action $B^{(0)}$. However, the additional correction due to the consideration of the self-consistent bounce, $B^{(2)}$, is 
of the same order of magnitude (about 50\%) as the purely one-loop correction $B^{(1)}$. This indicates that in some precision phenomenological studies when the conventional functional determinants are important, one may be required to consider the self-consistent bounce as well.   

Another motivation for considering the self-consistent bounce is to obtain a gauge-independent tunneling rate when gauge fields are involved. It is known that the effective action is gauge-dependent. By the Nielsen identity~\cite{Nielsen:1975fs,Fukuda:1975di,Plascencia:2015pga},
\begin{align}
    \left(\xi\frac{\partial \Gamma[\varphi;\xi]}{\partial \xi}+ K(\varphi;\xi) \frac{\delta\Gamma[\varphi;\xi]}{\delta\varphi} \right)=0\,,
\end{align}
where $\xi$ is the gauge parameter, the effective action is only gauge independent at its extrema. Then Eq.~\eqref{eq:decay3} gives a gauge-independent decay rate in principle only at the self-consistent bounce~\cite{Plascencia:2015pga}. Since the effective action must be truncated, this issue is usually more complicated. For more discussions related to this topic, see Refs.~\cite{Metaxas:1995ab,Garny:2012cg,Endo:2017gal,Endo:2017tsz,Gould:2021ccf,Lofgren:2021ogg,Hirvonen:2021zej}. 

Our method may find applications in physics of the early Universe (see, e.g., Refs.~\cite{Morrissey:2012db,Athron:2023xlk,Croon:2023zay}), where bubble nucleation can occur at finite temperature~\cite{Langer:1967ax,Langer:1969bc,Affleck:1980ac,Linde:1980tt,Linde:1981zj}.\footnote{Gravitational effects could also be very important in some situations~\cite{Coleman:1980aw,Gregory:2013hja,Canko:2017ebb,Mukaida:2017bgd,Hayashi:2020ocn,Shkerin:2021zbf,Shkerin:2021rhy,Gregory:2023kam}.} At high temperature, the phase transition is dominated by a static, i.e., Euclidean time independent, $O(3)$-symmetric bounce and the transition rate can be written as~\cite{Affleck:1980ac}
\begin{align}
    \gamma=\frac{\sqrt{|\lambda_-|}}{2\pi}\times 2|{\rm Im}\ln Z[\beta]| \,,
\end{align}
where $Z[\beta]$ is the thermal partition function with the temperature $T=1/\beta$, and $\lambda_-$ is the negative eigenvalue for the three-dimensional fluctuation operator at the classical bounce. The factor $\sqrt{|\lambda_-|}/(2\pi)$ above may be generalised to include the Langevin damping coefficient~\cite{Langer:1969bc,Hanggi:1990zz,Berera:2019uyp}. The calculation presented in this paper can be easily generalised to lower-dimensional bounce configurations, e.g., $O(3)$-symmetric static bounces that are relevant for thermal first-order phase transitions, and therefore can be applied to study radiative corrections in thermal first-order phase transitions~\cite{Ekstedt:2021kyx}. However, at finite temperatures, there are many more theoretical uncertainties~\cite{Croon:2020cgk,Guo:2021qcq}. The quantum corrections due to the self-consistent bounce might be detectable in experiments involving analogue systems in condensed matter~\cite{Braden:2017add, Braden:2019vsw,Jenkins:2023eez,Zenesini:2023afv}.

\section*{Acknowledgments}

The authors of this work are supported by the Engineering and Physical Sciences Research Council (grant No. EP/V002821/1). JA is also supported by the Leverhulme Trust (grant No. RPG-2021-299) and by the Science and Technology Facilities Council (grant No.  STFC-ST/X000753/1).

\newpage

\begin{appendix}
\renewcommand{\theequation}{\Alph{section}\arabic{equation}}
\setcounter{equation}{0}

\section{Solving Green's functions}
\label{app:Green}

In this appendix, we outline the procedure on solving Green's functions numerically. Suppose we want to solve the following Green's function
\begin{align}
    L[r] G(r,r')=\delta(r-r')\,,
\end{align}
where $L[r]$ is a second-order differential operator and $\{r,r'\}\in [0,\infty)$.

\subsection{\texorpdfstring{$G_l(r,r';\varphi)$}{TEXT} and \texorpdfstring{$\hat{G}_l(r,r')$}{TEXT}}

For $G_l(r,r';\varphi)$ and $\hat{G}_l(r,r')$, the boundary conditions are of the following type,
\begin{align}
    G(0,r')=0\,, \quad \left.\frac{\partial G(r,r')}{\partial r}\right|_{r=0}=0\,.
\end{align}
Assume that $\{f_1(r), f_2(r)\}$ are two independent (numerical) solutions to the equation $L[r] f(r)=0$. Then the Green's function can be written as
\begin{align}
    G(r,r')=\theta(r-r') [A_1^>(r') f_1(r) +A_2^>(r') f_2(r)]+\theta(r'-r) [A_1^<(r') f_1(r) +A_2^<(r') f_2(r)]\,,
\end{align}
where $A_1^{\gtrless}(r')$ and $A_2^{\gtrless}(r')$ shall be determined by the boundary conditions and matching conditions.

The matching conditions are from the following two requirements on $G(r,r')$: $(i)$ $G(r,r')$ is continuous at $r=r'$ when viewed as a function of $r$; $(ii)$ $\partial_r G(r,r')$ has a finite jump discontinuity at $r=r'$. The sign and magnitude of this jump depends on the coefficient of the second-derivative term in the operator $L[r]$. We will assume that this coefficient is $1$ to be consistent with the operators $\H_l[\varphi]$ and $\widehat{\H}_l$. Thus we have
\begin{subequations}
\begin{align}
    [A_1^>(r') f_1(r') +A_2^>(r') f_2(r')]-[A_1^<(r') f_1(r') +A_2^<(r') f_2(r')]=0\,,\\
    [A_1^>(r') f'_1(r') +A_2^>(r') f'_2(r')]-[A_1^<(r') f'_1(r') +A_2^<(r') f'_2(r')]=1\,.
\end{align}
\end{subequations}
After some algebra, one obtains
\begin{subequations}
\label{eq:relation-gtr-less}
\begin{align}
    A_1^{>}(r')-A_1^{<}(r')=-\frac{f_2(r')}{\W[f_1(r'),f_2(r')]}\,,\\
    A_2^{>}(r')-A_2^{<}(r')=\frac{f_1(r')}{\W[f_1(r'),f_2(r')]}\,,
\end{align}
\end{subequations}
where 
\begin{align}
    \W[f_1(r),f_2(r)]=f_1(r)f'_2(r)-f'_1(r) f_2(r)
\end{align}
is the Wronskian.

The functions $f_1$ and $f_2$ can be determined by the boundary conditions. We can choose, for example,\footnote{As a standard procedure for numerical calculations, one cannot take the boundary conditions strictly at $r=0$, but instead at a value $r_{\rm min}$ very close to zero. The boundary conditions at $r_{\rm min}$ can be obtained via the Taylor expansion $f(r_{\rm min})\approx f(0)+ f'(0) r_{\rm min}$ and $f'(r_{\rm min})=f'(0)+f''(0) r_{\rm rmin}$, where $f''(0)$ can be obtained by taking the limit $r\rightarrow 0$ for the equation to be solved. The infinity would be replaced by a large value $r_{\rm max}$ depending on the numerical bounce solution.}
\begin{align}
\label{eq:bdy-cond-f1-f2}
    f_1(0)=1\,, f_1'(r)|_{r=0}=0\,;\quad  f_2(r)|_{r\rightarrow\infty}\rightarrow 0\,, f'_2(r)|_{r\rightarrow\infty}\rightarrow 0\,.
\end{align}
In general, the boundary conditions for $f_2(r)$ lead to $f_2(0)\neq 0$, $f_2'(r)|_{r=0}\neq 0$. Then the boundary conditions for $G(r,r')$ require
\begin{align}
    A_2^{<}(r')=0\,,\quad  A^{<}_1(r')=0\,.
\end{align}
Substituting the above into Eqs.~\eqref{eq:relation-gtr-less}, one obtains $A_1^>(r')$ and $A_2^>(r')$ and thus also the Green's function
\begin{align}
\label{eq:G}
    G(r,r')=\theta(r-r')\left[-\frac{f_1(r) f_2(r')}{\W[f_1(r'),f_2(r')]}+\frac{f_1(r')f_2(r)}{\W[f_1(r'),f_2(r')]}\right]\,.
\end{align}
Apparently, $G(r,r)=0$.

For $\hat{G}_l(r,r')$, we have $L[r]=\widehat{\H}_l$. One can take
\begin{align}
\label{eq:f1-f2-M}
    f_1(r)=1\,,\qquad f_2(r)=\, \frac{K_{l+1}(m r)}{I_{l+1}(mr)}\,.
\end{align}
To see that $\widehat{\H}_l f_2(r)=0$, we recall that the operator $\widehat{\H}_l$ is obtained via the following procedure
\begin{align}
    \widehat{\M}_l \psi_l=0\quad \Rightarrow\quad \widehat{\H}_l\left(\frac{\psi_l}{\hat{\psi}_l}\right)=0\,,
\end{align}
where $\hat{\psi}_l \sim r^{-1} I_{l+1}(mr)$ (cf. Eq.~\eqref{eq:hat-psi-l}) is a solution to $\widehat{\M}_l \psi_l=0$. On the other hand, it is known that $\widehat{\M}_l \psi_l=0$ has another solution regular at $r\rightarrow\infty$, $r^{-1} K_{l+1}(mr)$. One immediately sees that $\widehat{\H}_lf_2(r)=0$.
With the basis solutions in Eq.~\eqref{eq:f1-f2-M}, we get Eq.~\eqref{eq:hatG-analytic-result}.

For $G_l(r,r';\varphi)$, we have $L[r]=\H[\varphi]$. In this case, to make the equation regular at $r\rightarrow\infty$ we can take 
\begin{align}
    f_2(r)=g_2(r) \frac{K_{l+1}(mr)}{I_{l+1}(mr)}\,,
\end{align} 
which gives 
\begin{align}
\frac{\d^2 g_2(r)}{\d r^2}-\left[2 m \frac{K_{l}(m r)}{K_{l+1}(m r)}+\frac{2l+1}{r}\right]\frac{\d g_2(r)}{\d r}-W(\varphi(r)) g_2(r)=0\,.   
\end{align}
The boundary conditions are $g_2(r)|_{r\rightarrow\infty}\rightarrow 1$, $\d g_2(r)/\d r|_{r\rightarrow\infty}\rightarrow 0$.

\subsection{\texorpdfstring{$\G(r,r')$}{TEXT}}

According to the boundary condition for $\Delta\varphi_b$, one has the boundary condition for the Green's function
\begin{align}
    \G(r=\infty,r')=0\,,\quad \left.\frac{\partial \G(r,r')}{\partial r}\right|_{r=0}=0\,.
\end{align}
In this case, we can choose two independent solutions $\{f_1(r), f_2(r)\}$ satisfying, e.g.,
\begin{align}
    f_1(0)=1\,, f_1'(r)|_{r=0}=0\,;\quad f_2(r)|_{r\rightarrow\infty}\rightarrow 0\,, f'_2(r)|_{r\rightarrow\infty}\rightarrow 1\,.
\end{align}
A similar analysis gives 
\begin{align}
    \G(r,r')=-\theta(r-r') \frac{f_1(r')f_2(r)}{\W[f_1(r'),f_2(r')]}-\theta(r'-r)\frac{f_2(r')f_1(r)}{\W[f_1(r'),f_2(r')]}\,,
\end{align}
Letting 
\begin{align}
    f_1(r)=g_1(r)\frac{I_{1}(m r)}{r}\,, \quad f_2(r)=g_2(r)\frac{K_{1}(m r)}{r}\,,
\end{align}
we have 
\begin{subequations}
\begin{align}
   &-g_1''(r)+\left[\frac{1}{r}-2 m \frac{I_{0}(m r)}{I_{1}(m r)}\right]g'_1(r)+W(\varphi_b^{(0)}(r))g_1(r)=0\,,  \\
   &-g_2''(r)+\left[\frac{1}{r}+2m \frac{K_{0}(m r)}{K_{1}(m r)}\right]g'_2(r)+W(\varphi^{(0)}_b(r)) g_2(r)=0\,,   
\end{align}
\end{subequations}
for which the boundary conditions can be chosen as, e.g., $g_1(0)=2/m$, $g'_1(r)|_{r=0}=0$ and $g_2(r)|_{r\rightarrow \infty}=0$, $g'_2(r)|_{r\rightarrow\infty}=1$. Note that $I_1 (m r)/r\rightarrow m/2$ for $r\rightarrow 0$.

\section{Renormalisation with dimensional regularisation}
\label{app:regularisation}

In this appendix, we rederive the expression~\eqref{eq:renormalisedD2} of the regularized one-loop functional determinant originally presented in Ref.~\cite{Baacke:2003uw}. 

We start with Eq.~\eqref{eq:S1-MSbar}. The divergent parts in $\Delta S$ are contained
in $\Delta S^{(2)}$,
\begin{align}
\label{eq:S2}
    \Delta S^{(2)}&=\frac{1}{2}\left[\ln\det (-\partial^2+m^2+W)-\ln\det(-\partial^2+m^2)\right]_{\O(W^2)}\,,
\end{align}
so that we can write
\begin{align}
    S_1= S_0^{\overline{\rm MS}}+\underbrace{\left[\Delta S-\Delta S^{(2)}\right]}_{\rm numerical\ part}+\underbrace{\left[\Delta S^{(2)}-(\Delta S)_{\rm pole}\right]}_{\rm analytical\ part}\,.
\end{align}
Our goal is to find two different expressions of $\Delta S^{(2)}$ such that the finite quantity $[\Delta S-\Delta S^{(2)}]$ can be computed numerically while the other finite quantity $[\Delta S^{(2)}-(\Delta S)_{\rm pole}]$ can be computed analytically. For this purpose, we assign the perturbation $W$ with a bookkeeping factor $\varepsilon$ and expand $\ln\det (-\partial^2+m^2+\varepsilon W)$ in $\varepsilon$. 

Numerically, we can write
\begin{align}
    \Delta S^{(2)}=\frac{1}{2}\sum_{l=0}^{\infty} (l+1)^2 \ln T_{l}(\infty;\varepsilon)|_{\O(\varepsilon^2)}\,.
\end{align}
where $T_l(\infty)$ is viewed as a function of $\varepsilon$ and expanded in powers of $\varepsilon$,
\begin{align}
    T_l(\infty;\varepsilon)=1+\varepsilon h_l^{(1)}(\infty)+\varepsilon^2 h_l^{(2)}(\infty)+\O(\varepsilon^3)\,. 
\end{align}
This gives
\begin{align}
    \ln T_l(\infty;\varepsilon)|_{\O(\varepsilon^2)}=h_l^{(1)}(\infty)+h_l^{(2)}(\infty)-\frac{1}{2} (h_l^{(1)}(\infty))^2\,.
\end{align}
To obtain $h_l^{(1)}(\infty)$ and $h_l^{(2)}(\infty)$, we substitute $T_l(r;\varepsilon)=1+\varepsilon h_l^{(1)}(r)+\varepsilon^2 h_l^{(2)}(r)+\O(\varepsilon^3) $ into Eq.~\eqref{eq:T_l}
\begin{align}
    \left[-\frac{\d^2}{\d r^2}-\left(2m\frac{I_{l+D/2-2}(m r)}{I_{l+D/2-1}(m r)}-\frac{2l+D-3}{r}\right)\frac{\d}{\d r} +\varepsilon W(\varphi(r))\right]T_l(r;\varphi)= 0\,,
\end{align}
where the perturbation term is associated with a factor of $\varepsilon$ now. Tracing the equations at the order of $\O(\varepsilon)$ and $\O(\varepsilon^2)$, we then obtain Eqs.~\eqref{eq:h1l-h2l} for $h_l^{(1)}(r)$ and $h_l^{(2)}(r)$. 

To obtain the analytical part $[\Delta S^{(2)}-(\Delta S)_{\rm pole}]$, we note that Eq.~\eqref{eq:S2} can be written as
\begin{align}
\label{eq:DeltaS2}
    \Delta S^{(2)}=\frac{1}{2} {\rm Tr}\left[(-\partial^2+m^2)^{-1}W\right]-\frac{1}{4} {\rm Tr}\left[(-\partial^2+m^2)^{-1} W(-\partial^2+m^2) W\right]\,.
\end{align}
The first term is 
\begin{align}
    \frac{1}{2} \int\d^4 x\,\langle x|(-\partial^2+m^2)^{-1} W(\hat{x})|x\rangle=\frac{1}{2} \left(\int\frac{\d^4 k}{(2\pi)^4} \frac{1}{k^2+m^2}\right)\int\d^4 x\, W(x)\,.
\end{align}
To do the $\overline{\rm MS}$ regularization, we perform the momentum integral in $d$-dimensional spacetime, using the useful formula
\begin{align}
    \mu^{4-d}\int\frac{\d^d k}{(2\pi)^d} \frac{1}{(k^2+m^2)^\alpha}=\mu^{4-d}\frac{1}{(4\pi)^{d/2}}\frac{\Gamma(\alpha-\frac{d}{2})}{\Gamma(\alpha)}\frac{1}{m^{2\alpha-d}}\,.
\end{align}
Expanding the integral result around $d=4$ (with $d=4-2\epsilon$), one then obtains the pole. Substracting the pole term according to the standard $\overline{\rm MS}$ rule, one then obtains $A_{\rm fin}^{(1)}$.

The second term on the RHS of Eq.~\eqref{eq:DeltaS2} can be written as
\begin{align}
   &-\frac{1}{4} \int\d^4 x\d^4 y\, W(x)W(y)\int\frac{\d^4 q}{(2\pi)^4}\e^{\i q(x-y)}\left[\int\frac{\d^4 k}{(2\pi)^4} \frac{1}{(k+q)^2+m^2}\frac{1}{k^2+m^2}\right]\notag\\
   =&-\frac{1}{4} \int\frac{\d^4 q}{(2\pi)^4} \widetilde{W}(q)\widetilde{W}(-q) \times \left[\int\frac{\d^4 k}{(2\pi)^4} \frac{1}{(k+q)^2+m^2}\frac{1}{k^2+m^2}\right]\,.
\end{align}
Again, the momentum integral inside the square brackets can be performed in $d$-dimensional spacetime by using the Feynman-parameter method, which in our case is
\begin{align}
\frac{1}{AB}=\int_0^1\d s\,\frac{1}{sA+(1-s)B}\,.
\end{align}
The final result of the momentum integral inside the square brackets  is
\begin{align}
    \frac{1}{16\pi^2}\int_0^1\d s\,\left[\frac{1}{\epsilon}-\gamma_{\rm E}+\ln 4\pi +\ln\frac{\mu^2}{s(1-s)q^2+m^2} \right]\,. 
\end{align}
Finally, subtracting the pole term ($1/\epsilon$ together with $-\gamma_{\rm E}+\ln 4\pi$) and doing the integral over $s$ gives $A_{\rm fin}^{(2)}$.

\end{appendix}

\bibliographystyle{utphys}
\bibliography{FVDref}{}

\end{document}